\documentclass{article}[14pt]
\usepackage{graphicx}
\usepackage{amsmath}
\usepackage{amssymb}
\usepackage{braket}
\usepackage{longtable}
\usepackage{mathtools}
\usepackage{hyperref}
\usepackage{breqn}
\usepackage{cite}

\usepackage{ifthen}
\usepackage{calc}
\usepackage{bm}
\usepackage{color}
\usepackage{psfrag}
\usepackage{wrapfig}
\usepackage{subfigure}
\usepackage{rotating}
\usepackage{tikz}
\usetikzlibrary{shapes.arrows}
\usetikzlibrary{decorations.pathmorphing}\usepackage{mdframed}
\usepackage{braket}

\numberwithin{equation}{section}
\newcommand\email[4]{#1@#2.#3.#4}

\newcommand{\be}{\begin{equation}}
\newcommand{\bs}{\begin{split}}
\newcommand{\ben}{\begin{equation}\nonumber}
\newcommand{\ee}{\end{equation}}
\newcommand{\es}{\end{split}}
\newcommand{\bi}{ \begin{itemize} }
\newcommand{\ei}{ \end{itemize} }

\begin{document}
	
	\title{Superconformal blocks for stress-tensor and chiral operators for 4D $\mathcal{N}=2$ superconformal field theories}
	\author{
		Subhadeep Rakshit\thanks{\email{srakshit}{sikkimuniversity}{ac}{in}}
		~~and~
		Subir Mukhopadhyay\thanks{\email{subirkm}{gmail}{com}{}}
	}
	\date{
		{\small Department of Physics, Sikkim University, 6th Mile, Gangtok 737102.}
	}
	
	\maketitle
	\begin{abstract}
		\small
		\noindent
		We have computed the superconformal partial wave for the mixed correlators involving $J$, $\phi$, and $\phi^\dagger$, where $J$ is the superconformal primary of 4D ${\mathcal N} = 2$ stress-tensor multiplet, $\phi$, and $\phi^\dagger$ are chiral and anti-chiral scalars, respectively in the channel  $\langle {J J}\phi\phi^{\dagger}\rangle$. We have used the superembedding formalism and computed the conformal partial wave corresponding to the exchange of long multiplet using the supershadow approach. This study provides part of the ingredients necessary for the study of conformal bootstrap for mixed correlator in 4D ${\mathcal N} = 2$ superconformal field theories.
	\end{abstract}
	\thispagestyle{empty}
	\clearpage

	\tableofcontents
	
	\section{Introduction}
	The conformal bootstrap program \cite{polyakov1974nonhamiltonian,mack1977duality,ferrara1973tensor} was introduced in two dimensions. Over the years, it has transpired that the analysis of conformal field theories using general consistency conditions from symmetries, unitarity, and associativity of the operator product expansion can impose substantial constraints on the spectrum of local operators of conformal field theories in any dimension. Recently, significant progress has been made for theories in dimensions higher than two \cite{poland2019conformal,rattazzi2008bounding}. In particular, it has been established that one can obtain the critical exponent of the 3D ising model using these methods \cite{el2014solving,kos2014bootstrapping}. 
	\paragraph{}
	
	As mentioned in \cite{beem2013n}, these methods are effective for theories characterised with global symmetries, and supersymmetric theories are natural candidates for the application of these methods.  In addition to the bosonic theories, the bootstrap method has been applied to the supersymmetric theories as well. In the latter, constraints from supersymmetry and protected aspects of the spectrum add to the power of its approach, leading to spectacular results.  Four dimensional superconformal theories for ${\mathcal N}=1,2,4$ have appeared in \cite{poland2011bounds,berkooz2014bounds,dolan2002superconformal,dolan2006conformal,dolan2004four,fortin2011current}, to list a few.  In this context, it may be mentioned that class $\mathcal{S}$ constructions \cite{gaiotto2012n,gaiotto2013wall} also lead to a number of SCFTs, and conformal bootstrap may help the classification of these theories.
	\paragraph{}
	
	One of the crucial ingredients in the applications of these techniques is four point correlators of the operators. The constraints arising from a single four point correlator are powerful enough to lead to bounds on scaling dimension, which turns out to be saturated for known CFT such as two dimensional minimal models \cite{beem2013n,Rychkov2009universal}. Four point correlators of the scalar operators in $\mathcal{N}=1$ SCFT in four dimensions have been studied in  \cite{fortin2011current,poland2011bounds,fitzpatrick2014covariant,khandker2014n,li2016most}. Mostly, they have focused on chiral operators and scalars in conserved current multiplets using different approaches. For $\mathcal{N}=2$ theories,  four point correlators of moment map operators and chiral operators were studied in \cite{beem2016n} and  \cite{lemos2016bootstrapping,cornagliotto2018bootstrapping}, where in the latter they considered Argyres-Douglas fixed point. 
	\paragraph{}
	
	The stress-tensor operator is universal in any local CFT, and four point correlators of operators of stress-tensor multiplet appeared in literature. In three dimension, the four point function of stress-tensor operator without the supersymmetry has been studied in \cite{dymarsky20183d}.  In four dimensions, the four point function of scalar operators of the stress-tensor multiplet for $\mathcal{N}=4$ theories was analysed in \cite{beem2013n,beem2017more}. They found that using bootstrap constraint a particular operator spectrum is singled out. It was conjectured that this extremal spectrum corresponds to $\mathcal{N}=4$ super Yang-Mills (SYM) at the S-dulity invariant value of $\tau = \theta + \frac{4\pi i}{g_{YM}^2}$. In this context, it may be mentioned that the bounds of scaling dimension and OPE coefficients for $\mathcal{N}=4$ SU(N) SYM was studied in \cite{chester2023level} for a wide range of N and $g_{YM}$ and they find the bounds are approximately saturated by weak coupling result at small $g_{YM}$, while for large N it interpolates between small $g_{YM}$ and strong coupling results.   In six dimensions, four point correlators of stress-tensor multiplet of $\mathcal{N}=(2,0)$  were studied\cite{beem20162}, which leads to various bounds. For the four dimensional $\mathcal{N}=2$ theories, the three point correlators of stress-tensor multiplets with arbitrary operators are discussed in  \cite{liendo2016stress,ramirez2016mixed} and the four point correlators of the scalar operators of stress- tensor multiplet were studied in \cite{li2020superconformal}. However, the bootstrapping has not been done yet.
	\paragraph{}
	
	Most of these works are focused on four point correlators of identical operators and bootstrapping eventually leads to interesting results. The study of correlators has also been extended to mixed correlators. As mentioned above, chiral operators and scalars in conserved current multiplets have been studied in $d=4, \mathcal{N}=1$. Bootstrapping of correlators of mixed correlators involving scalar chiral and real operators in $d=4, \mathcal{N}=1$ has been discussed in \cite{li2017bootstrapping}, which leads to bounds on the central charge and dimension of the real operator.  Mixed correlators in four dimension in ${\mathcal N} = 2$ superconformal theories have been studied involving moment map and chiral operators in \cite{Gimenez-Grau:2020jrx}. Similar study of chiral operators in four dimension for ${\mathcal N} = 2$ SCFT as an example of mixed correlator appeared in \cite{lemos2016bootstrapping}.
	\paragraph{}
	
	In the present work, we will consider mixed correlators in four dimensional $\mathcal{N}=2$ SCFTs involving scalars in the stress tensor multiplet and chiral multiplet. There are several works on the mixed correlators \cite{li2017bootstrapping,lemos2016bootstrapping,kos2014bootstrapping} to name a few. The power of mixed correlator in the context of three dimensional bootstrap has been elaborated in \cite{kos2014bootstrapping} and we expect this study will lead to a better understanding of $\mathcal{N}=2$ SCFTs. Analyzing the bootstrap of mixed correlators of chiral ($\Phi$), antichiral$(\Phi^\dagger$) and real ($R$) scalar multiplets, bounds on conformal dimension of $R$ and lower bound on central charge are obtained in \cite{li2017bootstrapping}.  Along the similar lines one would expect an analysis of the  bootstrap of the present results would lead to bounds on the scaling dimension of the operators and the OPE coefficients involved here.
	\paragraph{}

	
	Four point correlators of external operators in a conformal field theory can be expressed in terms of the conformal blocks, which correspond to the contribution due to the exchange of a primary operator and its descendants\cite{dolan2001conformal,dolan2004conformal}. In a non-supersymmetric CFT, conformal blocks for the external scalar operators are studied in 
	\cite{dolan2001conformal,dolan2004conformal} and are expressed in terms of conformal invariant cross-ratios. An account of various methods used to compute conformal blocks can be obtained in \cite{poland2019conformal}.
	\paragraph{}
	
	Once we move to superconformal field theories, the three point functions of primary operators residing in the same supersymmetry multiplet are related to each other through superconformal algebra. One can construct the superconformal blocks, which sum up contributions from all the operators in a given supersymmetry multiplet. Therefore, the superconformal blocks are composed of several conformal 
	blocks and the explicit form depends on the particular superconformal algebra and the representations. For $\mathcal{N}=1$ SCFTs, superconformal block decompositions have been discussed in \cite{poland2011bounds,fortin2011current,li2017bootstrapping}, in the context of chiral and antichiral external operators. However, this method becomes quite cumbersome for ${\mathcal N}=2$ or higher. 
	\paragraph{}
	
	There are several approaches to constructing superconformal blocks. In the case of half-BPS external operators for extended supersymmetry, one can use the superconformal Ward identities to obtain the superconformal blocks \cite{nirschl2005superconformal,dolan2002superconformal,dolan2004four}.
	\paragraph{}
	
	A more general approach is to use the Casimir operator of the superconformal group, which can be applied in any dimension. One can identify conformal partial waves (which are the same as conformal blocks up to certain kinematic factors) as eigenfunctions of quadratic Casimir operators belonging to the eigenvalue related to the quantum number of the representation. The superconformal blocks for the chiral operators have been computed in \cite{fitzpatrick2014covariant} using this method. This method has also been used to compute the superconformal blocks (in theories with eight supercharges) for external protected scalar operators, which are the lowest component of the flavour current multiplet \cite{bobev2015bootstrapping,bobev2017superconformal,fitzpatrick2014covariant}. However, for extended supersymmetry, it involves more than one nilpotent superconformal invariant, and this approach becomes cumbersome.

	\paragraph{}
	
	In this work, we have used the shadow approach to compute superconformal partial waves. This involves a supersymmetric generalization of the embedding formalism, in which the conformal transformations are realized linearly.  In this approach, one describes the superembedding space in terms of supertwisters, which transform naturally under the superconformal group. The superfields can be lifted to the superembedding space. Then one writes a manifestly invariant projector onto an irreducible representation of the superconformal group. The partial waves can then be written as manifestly invariant supertwistor integrals. This approach has been developed in \cite{fitzpatrick2014covariant,khandker2012superembedding} and the superconformal blocks for the general external operator have been computed in \cite{khandker2012superembedding,li2017bootstrapping}. The relation between the superembedding space and the superspace has been studied in \cite{goldberger2012superembedding,goldberger2013superembedding,maio2012superembedding, kuzenko2000correlation}. In the superspace formalism, study of 4D $\mathcal{N}=2$ have been carried out \cite{liendo2016stress,Park:1999pd,kuzenko2000correlation,arutyunov2001non}.
	\paragraph{}
	
	We have organized the paper as follows. In Section 2 we review the $4D$ $\mathcal{N}=2$ superconformal algebra, its representation, and the OPE selection rules for $\mathcal{J}\times\mathcal{J}$ and $\Phi\times\Phi^\dagger$, where $\mathcal{J}$, $\Phi$ and $\Phi^\dagger$ are the superfields representing the stress-tensor, chiral and anti-chiral scalar multiplets respectively . In the next section, we discuss the superembedding space formalism with $\mathcal{N}=2$ supersymmetry, the superconformal invariants, and the tensor structures. In Section 4 we discuss the three point function of $\langle\mathcal{J}\mathcal{J}\mathcal{O}\rangle$ and $\langle \Phi\Phi^{\dagger}\mathcal{O}\rangle$ where $\mathcal{O}$ is the long multiplet. We computed the superconformal partial wave for the four point correlation function $\langle {J J}\phi\phi^{\dagger}\rangle$ in Section 5. We discuss our results with reference to the decomposition of $\mathcal{N}=2$ supermultiplet into several $\mathcal{N}=1$ multiplets in Section 6, which provides us the nontrivial consistency check for odd and even $l$. In Section 7 we conclude. Section A and Section B are a small sampling of computations of superconformal invariants and conformal invariants.
	
	\section{Revisit  $\mathcal{N}=2$ superconformal algebra and selection rule for OPEs}
	
	\subsection{Elements of  4D $\mathcal{N}=2$ superconformal algebra}
	
	In this subsection, we review the essential ingredients of 4D $\mathcal{N}=2$ superconformal algebra following \cite{dolan2003short, liendo2016stress,li2020superconformal}. The bosonic conformal algebra is generated by the generators $\left\{\mathcal{P}_{\alpha,\dot{\alpha}},\mathcal{K}^{\alpha,\dot{\alpha}},\mathcal{M}_{\alpha}^{\beta},\bar{\mathcal{M}}_{\dot{\beta}}^{\dot{\alpha}},D\right\}$. Unless otherwise mentioned, we will be using  $\alpha=\pm$ and $\dot{\alpha}=\pm$ as the Lorentz indices. The conformal group is given by $SU(2,2)\sim SO(4,2)$.
	
For the extended supersymmetries, such as, the $\mathcal{N}=2$ superconformal algebra is further augmented by including the fermionic Poincar\'{e} and conformal supercharges, $\left\{\mathcal{Q}^i_{\alpha},\bar{\mathcal{Q}}_{i\dot{\alpha}},\mathcal{S}^{\alpha}_{i},\bar{\mathcal{S}}^{i\dot{\alpha}}\right\}$. The conformal group $SU(2,2)$ is also extended into its supersymmetric extension $SU(2,2|2)$. In addition, they are characterised by R-symmetry generators, and for $\mathcal{N}=2$ the R-symmery group is given by $SU(2)_R\times U(1)_r$ and we will denote the generators by $\left\{\mathcal{R}^i_{j}, r\right\}$ with $SU(2)_R$ indices $i,j =1,2$.
	\paragraph{}
	
	The representation of the superconformal group $SU(2,2|2)$ of a general multiplet can be obtained from the highest weight or superconformal primary. The highest weight states, by definition, are annihilated by the conformal supercharges $({\mathcal S}, \bar{\mathcal S})$. They are characterised by the Dynkin labels of the conformal group and the R-symmetry groups, which are given by the quantum numbers $(\Delta,j,\bar{j}, R,r)$, where $(\Delta,j,\bar{j})$ and $(R,r)$ are the Dynkin labels of the conformal group and the R-symmetry group, respectively. The other states in the multiplet are generated by the action of the Poincar\'{e} supercharges.

General supermultiplets  denoted by $\mathcal{A}^{\Delta}_{R,r,(j,\bar{j})}$ (we follow the convention of \cite{dolan2003short}) is obtained by application of $Q^i_{~\alpha}, \bar{Q}_{i~\dot\alpha}$. Unitarity imposes bounds on the conformal dimension of the multiplets, which is known as unitarity bounds. For general long multiplet, the unitarity bound is given by,
	\begin{equation}
		\Delta\geq 2+2j +2R  + r ,
	 \hspace{0.2cm} 2 + 2\bar{j} + 2 R - r.	\end{equation}

	If the highest weight is annihilated by some combination of $\mathcal{Q}$ and $\mathcal{\bar Q}$ the supermultiplet undergoes shortening. There are several kinds of shortening conditions \cite{dolan2003short,kinney2007index} depending on the Lorentz and $SU(2)_R$ quantum numbers of the charges that annihilate the highest weight. 
	
	If $j=0$ we may impose 
	\be Q^1_{~\alpha} \ket{R,r}^{hw} = 0,	\ee which requires $\Delta = 2 R + r$. This multiplet is denoted by ${\mathcal B}_{R,r(0,\bar{j})}$. Similarly, for ${\bar j}=0$  we may impose \be \bar{Q}_{2\dot\alpha} \ket{R,r}^{hw} = 0,	\ee which requires $\Delta = 2 R - r$, and the corresponding multiplet is denoted by $\bar{\mathcal B}_{R,r(j,0)}$.
	
	For $R=0$, when the state is annihilated by $Q^i_{~\alpha}$ for both $i=1,2$, the multiplet is generated by the action of $\bar{Q}$ and it leads to a chiral multiplet ${\mathcal E}_{r(0,\bar{j})}$. Similarly, there is a corresponding conjugate $\bar{\mathcal E}_{r(j,0)}$. Though representation theory allows such ${\mathcal N}=2$ chiral multiplets with non-zero $\bar{j}$ and $j$ respectively, but such exotic multiplets do not occur in any known ${\mathcal N}=2$ SCFT \cite{buican2014constraints}. 
	In fact, it has been shown in \cite{manenti2020differential} that the exotic chiral primaries cannot appear in any local SCFT. They begin by constructing a 3 point function involving exotic chiral primary, its conjugate, and stress tensor multiplet and show that it does not satisfy Ward identity unless $j = 0$.
	The scalar ${\mathcal E}_r := {\mathcal E}_{r(0,0)}$ multiplets are the half-BPS multiplets of Coulomb type.
	
	On the other hand, imposing $ Q^1_{~\alpha} \ket{R,r}^{hw} =0$ and $ \bar{Q}_{2\dot\alpha} \ket{R,r}^{hw} = 0$ simultaneously, requires $r=0$ and $j=\bar{j}=0$, leading to a short supermultiplet  $\hat{\mathcal B}_R$. These scalar multiplets are half-BPS multiplets of Higgs type.
	
	The other shortening conditions are imposed as follows. For $j > 0$ one can impose
	\be \epsilon^{\alpha\beta}\mathcal{Q}^i_{\beta}\ket{R,r}^{hw}_\alpha = 0 \ee while  for $j=0$ this condition is replaced by 
	\be (\mathcal{Q}^i)^2 \ket{R,r}^{hw} = 0, \ee
	where $(\mathcal{Q}^i)^2= \epsilon^{\alpha\beta}\mathcal{Q}^i_{\alpha}\mathcal{Q}^i_{\beta} $ (no sum on $i$).
	If we require this only for $i=1$ we get $\Delta=2+2j+2R+r$ and $\Delta=2+2R+r$ respectively.  This leads to the supermultiplet ${\mathcal C}_{R,r(j,\bar{j})}$. 
	
Similarly, the conjugate conditions are 
\be\begin{split} 
\epsilon^{\dot\alpha\dot\beta}\bar{\mathcal{Q}}_{i \dot\beta}\ket{R,r}^{hw}_{\dot\alpha} &= 0 , \quad \bar{j}\ge 0, \\
 (\bar{\mathcal Q}_i)^2 \ket{R,r}^{hw} & = 0,\quad \bar{j}=0,
 \end{split} \ee
 and this multiplet is denoted by $\bar{\mathcal C}_{R,r(j,\bar{j})}$.
 
 ${\mathcal C}_{R,r(j,\bar{j})}, \bar{\mathcal C}_{R,r(j,\bar{j})}$ are called semi-short multiplets. If the semi-shortening condition is applied for both $Q$ and $\bar{Q}$, we obtain a semi-short multiplet $\hat{\mathcal C}_{R,(j,\bar{j})}$, with $r=\bar{j}-j$.

  We have included a table \ref{tab:1} from \cite{liendo2016stress,li2020superconformal} where the unitary irreducible representation of the $\mathcal{N}=2$ superconformal algebra has been shown. Though we will be discussing only a few multiplets, that are relevant for our purpose, we have included a full list of multiplets in the table. It may be noted that for the ${\mathcal C}$ multiplets, $\Delta$ satisfies the bounds imposed for the long multiplets, which are saturated for the multiplet $\hat{\mathcal C}_{R,(j,\bar{j})}$.
	\begin{table}	\centering
		\begin{tabular}{|c|c|c|} 
			\hline
			Shortening & Quantum Number Relations & Multiplet \\
			\hline\hline
			$\oslash$ & $\Delta\geq 2+2j +2R  + r, 2 + 2\bar{ j} + 2 R  - r $\hspace{5 cm} & $\mathcal{A}^\Delta_{R,r, (j , \bar{j})}$\\
			\hline	
			$\mathcal{B}^1$ & $\Delta = 2 R + r $ \hspace {5 cm} $j=0$&$\mathcal{B}_{R,r(0,\bar{j})}$\\
			
			\hline
			
			$\bar{\mathcal{B}}_2$ &$\Delta=2R-r$ \hspace{5 cm} $\bar{j}=0$& $\bar{\mathcal{B}}_{R,r(j,0)}$\\
			
			\hline
			$\mathcal{B}^1\cap \mathcal{B}^2$ & $\Delta = r $ \hspace{5.9 cm} $R=0$ & $\mathcal{E}_{r(0,\bar{j})}$ \\
			\hline
			$\bar{\mathcal{B}}_1\cap \bar{\mathcal{B}}_2$ & $\Delta = - r $ \hspace{5.6 cm} $R=0 $ & $\bar{\mathcal{E}}_{r(j,0)}$ \\
			\hline
			$\mathcal{B}^1 \cap \bar{\mathcal{B}}_2$ & $\Delta  = 2 R $ \hspace{4.4 cm} $ j= \bar{j} = r = 0 $ & $\hat{\mathcal{B}}_R$ \\
			\hline\hline
			$\mathcal{C}^1$ & $\Delta = 2 + 2 j + 2 R + r $\hspace{4.6cm} &$\mathcal{C}_{ R , r ( j , \bar{j} )}$\\
			\hline
			$\bar{\mathcal{C}}_2$ & $\Delta = 2 + 2 \bar{j} + 2 R - r $\hspace{4.6cm} &$\bar{\mathcal{C}}_{R,r(j , \bar{j} ) }$\\
			\hline
			$\mathcal{C}^1\cap \mathcal{C}^2$ & $\Delta = 2 + 2 j + r $ \hspace{4.3 cm} $R=0$& $\mathcal{C}_{0,r( j , \bar{j} )}$ \\ 
		\hline
			$\bar{\mathcal{C}}_1\cap \bar{\mathcal{C}}_2$ & $\Delta = 2 + 2 \bar{j} - r $ \hspace{4.3 cm} $R=0$& $\bar{\mathcal{C}}_{0,r(j , \bar{j})}$ \\
			\hline
			$\mathcal{C}^1\cap \bar{\mathcal{C}}_2$ & $\Delta = 2 + 2 R + j + \bar{j} $ \hspace{2.7 cm} $r = \bar{j} - j $ & $\hat{\mathcal{C}}_{R(j , \bar{j})}$ \\
			\hline\hline
			$\mathcal{B}^1\cap \bar{\mathcal{C}}_2$ & $\Delta = 1 + \bar{j} + 2 R $ \hspace{3.5 cm} $ r= \bar{j} + 1 $ & $\mathcal{D}_{R(0,\bar{j})}$ \\
			\hline
			$\bar{\mathcal{B}}_2\cap \mathcal{C}^1$ & $\Delta=1 + j + 2 R $ \hspace{3.2 cm} $-r = j +1 $& $\bar{\mathcal{D}}_{R(j ,0)}$ \\
			\hline
			$\mathcal{B}^1\cap \mathcal{B}^2\cap\bar{\mathcal{C}}_2$ & $\Delta=r=1+\bar{j} $ \hspace{4.3 cm} $R=0$& $\mathcal{D}_{0(0, \bar{j})}$ \\
			\hline
			$\mathcal{C}^1\cap \bar{\mathcal{B}}_1\cap\bar{\mathcal{B}}_2$ & $\Delta=-r=1+j$ \hspace{4.0 cm} $R=0$& $\bar{\mathcal{D}}_{0(j,0)}$ \\
			\hline
		\end{tabular}
		\caption{Classification of the unitary irreducible representations of the $\mathcal{N}=2$ superconformal algebra}
		\label{tab:1}
	\end{table}

	\subsection{ Three point functions in $\mathcal{N}=2$ superspace}
	
	\textbf{Three point function $\langle\mathcal{J}\mathcal{J}\mathcal{O}\rangle$ and  OPE selection rules:}
	
	The three point functions consisting of two identical stress tensor primary operators $J$ and a third arbitrary operator have been discussed in \cite{li2020superconformal,liendo2016stress}. In particular the three point function  $\langle\mathcal{J}\mathcal{J}\mathcal{J}\rangle$ has been discussed in \cite{kuzenko2000correlation}. In this section, we briefly review the three point function consisting of two stress tensor primary operators $J$ and a third arbitrary operator in terms of variables of superspace. 
	
	In this subsection, we will discuss the three point functions in terms of variables of superspace. We will begin with the three point function consisting of two identical stress tensor primary operators $J$ and a third arbitrary operator. These have been discussed in \cite{li2020superconformal,liendo2016stress} and we will briefly review it. 
	
	The superfield representing the stress-tensor multiplet is given by,
	\begin{equation}
		\mathcal{J}(x,\theta,\bar{\theta})=J(x)+J^{i}_{\alpha\dot{\alpha}j}\theta^{\alpha}_{i}\bar{\theta}^{j\dot{\alpha}}+....    .
	\end{equation}
	This superfield satisfies the following reality conditions and the conservation equations
	\begin{equation}
		\mathcal{J}=\bar{\mathcal{J}} , \qquad D^{\alpha i}D^{j}_{\alpha}\mathcal{J}=0, \qquad \bar{D}^i_{\dot{\alpha}}\bar{D}^{j\dot{\alpha}}\mathcal{J}=0,
	\end{equation}
	where $D^{\alpha i}$ and $\bar{D}^{j\dot{\alpha}}$ are the ${\mathcal N}=2$ covariant derivative  \cite{liendo2016stress}.
	In order to discuss $\mathcal{N}=2$ SCFT correlators, we introduce superconformal covariant coordinates in superspace following \cite{liendo2016stress}, $z^I=(\tilde{x}^{\dot{\alpha} \alpha},\theta_{i},\bar{\theta}^i)$. Following \cite{liendo2016stress} we use $x_{\alpha\dot{\alpha}}=x^{\mu}(\sigma_{\mu})_{\alpha\dot{\alpha}}$ and $\tilde{x}^{\dot{\alpha}\alpha}=x^{\mu}(\tilde{\sigma}_{\mu})^{\dot{\alpha}\alpha}$ and the chiral combinations are given by $\tilde{x}^{\dot{\alpha}\alpha}_{\pm}=\tilde{x}^{\dot{\alpha}\alpha}\mp2i\theta^{\alpha}_{i}\bar{\theta}^{\dot{\alpha}i}$. 
	
	Using two such points $(z_1,z_2)$ one can construct the variables transform as a product of two tensors,
	\begin{equation}	\tilde{x}^{\dot{\alpha}\alpha}_{\bar{1}2}=\tilde{x}^{\dot{\alpha}\alpha}_{1-}-\tilde{x}^{\dot{\alpha}\alpha}_{2+} -4i \theta^{\alpha}_{2 i}\bar{\theta}^{\dot{\alpha} i}_1=(\tilde{x}_{12})_{-}^{\dot{\alpha}\alpha} ,
		\qquad
		z^I_{12}=z^I_1-z^I_2
	\end{equation}	
	
The superconformally covariant coordinates $Z_3=(X_3,\Theta_3,\bar{\Theta}_3)$ 	are constructed using three points $(z_1, z_2, z_3)$ as follows,
	\begin{equation}
		X_{3\alpha \dot{\alpha}}=\frac{x_{3\bar{1}\alpha\dot{\beta}}\tilde{x}^{\dot{\beta}\beta}_{\bar{1}2}x_{2\bar{3}\beta\dot{\alpha}}}{(x_{3\bar{1}})^{2}(x_{2\bar{3}})^{2}}, \qquad	\bar{X}_{3\alpha\dot{\alpha}}=X^{\dagger}_{3\alpha\dot{\alpha}}=-\frac{x_{3\bar{2}\alpha\dot{\beta}}\tilde{x}_{\bar{2}1}^{\dot{\beta}\beta}x_{1\bar{3}\beta\dot{\alpha}}}{(x_{3\bar{2}})^{2}(x_{1\bar{3}})^{2}}
	\end{equation}
	\begin{equation}
		\Theta^{i}_{3\alpha}=i(\frac{x_{\bar{2}3\alpha\dot{\alpha}}}{x^2_{\bar{2}3}}\bar{\theta}_{32}^{\dot\alpha i}-\frac{x_{\bar{1}3\alpha\dot{\alpha}}}{x^2_{\bar{1}3}}\bar{\theta}^{\dot\alpha i}_{31}),\qquad \bar\Theta_{3\dot{\alpha}i}=i(\theta^{\alpha}_{32 i}\frac{x_{\bar{3}2\alpha\dot\alpha}}{x^2_{\bar{3}2}}-\theta^{\alpha}_{31 i}\frac{x_{\bar{3}1}\alpha\dot{\alpha}}{x^2_{\bar{3}1}}).
	\end{equation}
	By cyclically permutating $z_1$, $z_2$ and $z_3$ one can construct $Z_1$ and $Z_2$ in a similar manner.
	It follows that $\bar{X}$ and $X$ are related through the following relation
	\begin{equation}
		\bar{X}_{3\alpha\dot{\alpha}}=X_{3\alpha\dot{\alpha}}-4i\Theta_{3{\alpha}}^i\bar\Theta_{3\dot{\alpha}i}.
	\end{equation}
	
	We also have
	\begin{equation}
		X_3^2 = \frac{x_{\bar{1}2}}{x_{\bar{1}3}^2 x_{\bar{3}2}^2}, \qquad 
		\bar{X}_3^2 = \frac{x_{\bar{2}1}}{x_{\bar{3}1}^2 x_{\bar{2}3}^2}
	\end{equation}
	
	Using $Z_1$, $Z_2$ and $Z_3$ one can obtain a superconformal invariant given by
	\begin{equation}
		{\bf u} = \frac{X_1^2}{\bar{X}_1^2} = \frac{X_2^2}{\bar{X}_2^2}=\frac{X_3^2}{\bar{X}_3^2}.
	\end{equation}
	This variable appears only in the superconformal theories and is absent in non-supersymmetric conformal theories \cite{goldberger2012superembedding}. In addition, ${\mathcal N}=2$ theories involve one more superconformal invariant  given by
	\begin{equation}
		-{\bf w}^\prime = \frac{X_1 . \bar{X}_1}{\sqrt{ X_1^2 \bar{X}_1^2}}.
	\end{equation}
	
	
	In $\mathcal{N}=2$ superspace, the three point function can be written as
	\begin{equation} \langle\mathcal{J}(z_1)\mathcal{J}(z_2)\mathcal{O}(z_3)\rangle=\frac{1}{(x_{\bar{1} 3})^2(x_{\bar {3} 1})^2(x_{\bar{2}3})^2(x_{\bar{3}2})^2} H(Z_3).
	\end{equation}
	
	The function $H$ satisfies the scaling condition,
	\begin{equation}
		H({\lambda\bar{\lambda}X_3,\lambda\Theta_3,\bar{\lambda}\bar{\Theta}_3})=\lambda^{2a}\bar{\lambda}^{2\bar{a}}H(X_3,\Theta_3,\bar{\Theta}_3)
	\end{equation} 
	with $a-2\bar{a} = 2-q$ and $\bar{a}-2 a = 2-q$, where $\Delta = q + \bar{q}$ and $r=q-\bar{q}$.
	
	An additional restriction is being imposed on $H$ coming from the conservation equations of $\mathcal{J}$. These imply,
	\begin{equation}
		\frac{\partial^2}{{\partial\Theta}_{3{\alpha}}^i\partial{\Theta}^{\alpha j}_{3}}H(Z_3)=0,\qquad \frac{\partial^2}{\partial\bar{\Theta}_{3i}^{\dot{\alpha}}\partial\bar{\Theta}_{3\alpha j}}H(Z_3)=0,
	\end{equation}
	\begin{equation}
		{\mathcal D}^{\alpha}_{i}{\mathcal D}_{\alpha j}H(Z_3)=0,\qquad \tilde{{\mathcal D}}^{\dot{\alpha}i}\tilde{{\mathcal D}}^{j}_{\dot\alpha}H(Z_3)=0
	\end{equation}
	where 
	\begin{equation}
		{\mathcal D}_{\alpha i}=\frac{\partial}{\partial\Theta_{3}^{\alpha i}}+4i\bar{\Theta}^{\dot{\alpha}}_{3i}\frac{\partial}{\partial X_{3}^{\dot{\alpha}\alpha}},
		\qquad \tilde{{\mathcal D}}^{\dot{\alpha i}}=\frac{\partial}{\partial \bar{\Theta}_{3\dot{\alpha}i}}-4i\Theta^{i}_{3\alpha}\frac{\partial}{\partial X_{3\alpha\dot{\alpha}}}.	\end{equation}
	
	The correlator also needs to be invariant under $z_1 \leftrightarrow z_2$, which implies $(X_3, \Theta_3, \bar{\Theta}_3) \leftrightarrow (-\bar{X}_3, - \Theta_3, -\bar{\Theta}_3)$ 
	Using the above restrictions, one can write down expressions of the three point correlators for various operators $\mathcal{O}$. In particular, the three point function $\langle\mathcal{J}(z_1)\mathcal{J}(z_2)\mathcal{J}(z_3)\rangle$ has been written by \cite{kuzenko2000correlation} .
	
	
	 Next, we need to consider the operators that can appear in the $\mathcal{J}\times\mathcal{J}$ operator product expansion from the selection rule. All possible multiplets $\mathcal{O}$ that appear in $\mathcal{J}\times\mathcal{J}$ OPE have been studied by Liendo et.al.\cite{liendo2016stress}. The stress-tensor multiplet is denoted by $\hat{\mathcal{C}}_{0(0,0)}$, which contains spin two conserved current\cite{li2020superconformal,liendo2016stress} and the selection rule of the OPE is 
	\begin{equation}\label{eq:19}
		\begin{split}
		\hat{\mathcal{C}}_{0(0,0)}\times\hat{\mathcal{C}}_{0(0,0)}\sim\mathcal{I}+\hat{\mathcal{C}}_{0(\frac{l}{2},\frac{l}{2})}+\hat{\mathcal{C}}_{1(\frac{l}{2},\frac{l}{2})}+\mathcal{C}_{\frac{1}{2},\frac{3}{2}(\frac{l}{2},\frac{l+1}{2})}+\\ \mathcal{C}_{0,3(\frac{l}{2},\frac{l+2}{2})}+\mathcal{C}_{0,0(\frac{l+2}{2},\frac{l}{2})}+\mathcal{C}_{0,0(\frac{l+4}{2},\frac{l}{2})}+\mathcal{A}^{\Delta}_{0,0(\frac{l}{2},\frac{l}{2})}+\mathcal{A}^{\Delta}_{0,0(\frac{l+2}{2},\frac{l}{2})}+\mathcal{A}^{\Delta}_{0,0(\frac{l+4}{2},\frac{l}{2})}+c.c. ..
		\end{split}
	\end{equation}
	
	The three point correlators have been studied in \cite{li2020superconformal} for the long multiplets $\mathcal{A}^{\Delta}_{0,0(\frac{l}{2},\frac{l}{2})}$, $\mathcal{A}^{\Delta}_{0,0(\frac{l+2}{2},\frac{l}{2})}$ and $\mathcal{A}^{\Delta}_{0,0(\frac{l+4}{2},\frac{l}{2})}$. 
	\\
	\\
	
	\noindent \textbf{Three point function $\langle \Phi\Phi^\dagger \mathcal{O}\rangle $ and OPE selection rules:}
	\\
	
	Similar considerations can be used to determine the second three point function of our interest, which is given by $\langle\Phi\Phi^\dagger\mathcal{O}\rangle$. 
	It takes the following form in superspace covariant coordinates \cite{poland2011bounds}
	\begin{equation}
		\langle\Phi(z_1)\Phi^{\dagger}(z_2)\mathcal{O}(z_3)\rangle\propto\frac{1}{x_{\bar{3}1}^{2q}x^{2q}_{\bar{2}3}}t(X_3,\Theta_{3},\bar{\Theta}_{3})
	\end{equation}
	In this case, the chirality condition will impose the additional constraints \cite{poland2011bounds}. The chirality condition $\bar{D}^{\dot{\alpha}}_1t=0$ implies $t$ is a function of $\bar{X}_3$ and $\bar{\Theta}_3$ and other constraint will come from $D_2^{\alpha}$, which reads
	\begin{equation}
		D_2^{\alpha}t(\bar{X}_3,\bar{\Theta}_3)=i\frac{x_{2\bar{3}}^{\dot{\alpha}\alpha}}{x^2_{\bar{2}3}}\frac{\partial}{\partial\bar{\Theta}_3^{\dot{\alpha}}}t(\bar{X}_3,\bar{\Theta}_3)=0
	\end{equation}
	From the above, we can conclude that the three point function $\langle\Phi(z_1)\Phi^{\dagger}(z_2)\mathcal{O}(z_3)\rangle$ depends only on $\bar{X}_3$. The scaling conditions further fix it to,
	\begin{equation}\label{eq:20}
		\langle\Phi(z_1)\Phi^{\dagger}(z_2)\mathcal{O}^{(l)}(z_3)\rangle\propto\frac{1}{\langle\bar{3}1\rangle^{q}\langle\bar{2}3\rangle^{q}}{\bar{X}^{\Delta_{\mathcal{O}}-2\Delta_{\Phi}-l}_{3}\bar{X}_3^{\mu_1}...\bar{X}_3^{\mu_l}}	\end{equation}
	
	The chiral and anti-chiral operators are the scalars of $\mathcal{E}_r$ multiplets \cite{beem2016n}. The operator $\mathcal{O}(x_3)$ appears in their OPE if they are $SU(2)_R$ singlet, $r_{\mathcal{O}}=0$ and has spin $  \bar{j} =j$. The OPE selection rule is as follows \cite{beem2016n},
	\begin{equation} \label{eq:21}
		\mathcal{E}_{r_0}\times\bar{\mathcal{E}}_{-r_0}\sim\mathcal{I}+\hat{\mathcal{C}}_{0,(\frac{l}{2},\frac{l}{2})}+\mathcal{A}^{\Delta}_{0,0(\frac{l}{2},\frac{l}{2})}
	\end{equation}
	
	\paragraph{}
	
	Comparing the selection rules for OPE of $\mathcal{J}\times\mathcal{J}$ and $\Phi\times\Phi^{\dagger}$ as given in (\ref{eq:19}) and (\ref{eq:21}) we find the common multiplets are $\hat{\mathcal{C}}_{0(\frac{l}{2},\frac{l}{2})}$, $\mathcal{A}^{\Delta}_{0,0(\frac{l}{2},\frac{l}{2})}$ and identity. 
	
	Let us consider the multiplet $\hat{\mathcal{C}}_{0(j,\bar{j})}$ which is a special version of the semi-short multiplet $\hat{\mathcal{C}}_{R(j,\bar{j})}$ and can be obtained by setting the $R$ charge $0$. The multiplet $\hat{\mathcal{C}}_{0(j,\bar{j})}$ obeys the shortening condition $\mathcal{C}^1\cap\mathcal{C}^2\cap\bar{\mathcal{C}}_1\cap\bar{\mathcal{C}}_2$ and it contains conserved higher spin currents, which involve spin greater than 2, and the stress tensor multiplet does not contain such currents. A single higher spin current indicates the existence of an infinite number of high spin currents. As explained in \cite{Alba:2013yda,maldacena2013constraining} the existence of this multiplet implies that the theory is a free theory. So in the case of interacting theory, we need not consider the multiplet $\hat{\mathcal{C}}_{0(j,\bar{j})}$, except ${\mathcal J}$.
	\paragraph{}
	
	\section{Review of superembedding space formalism}
	
	\subsection{Superembedding space} In superembedding space, the coordinates transform linearly under superconformal transformations. The local operators of a conformal field theory can be uplifted to respective superembedding space, and in terms of them, one can compute the superconformal correlators.
	Superembbeding space formalism has been discussed in this context    \cite{kuzenko2022supertwistor,siegel2012embedding,fitzpatrick2014covariant,goldberger2012superembedding,Alba:2015upa,khandker2014n
}. For our purpose, we will review superembedding formalism for $\mathcal{N}=2$ SCFT in this section. 
	In this work, we have used the notation and convention of \cite{fitzpatrick2014covariant,khandker2014n}. 
	\paragraph{}
	
	The basic building blocks of the superembedding space are supertwistors
	\begin{equation}
		\mathbb{Z}_A=\begin{pmatrix}
			Z_\alpha\\
			Z^{\dot{\alpha}}\\
			Z_i\end{pmatrix}\in \mathbb{C}^{4|\mathcal{N}},
	\end{equation}
	Here $Z_{\alpha}$,$Z^{\dot{\alpha}}$ are the four bosonic components, and $Z_i$ are the $\mathcal{N}$ fermionic components.
	
	 The superconformal group $SU(2,2|\mathcal{N})$ is the subgroup of $SL(4|\mathcal{N})$ that preserves the dot product 
	\begin{equation}
		\langle\mathbb{Z}_1, \mathbb{Z}_2\rangle=\mathbb{Z}_1^{\dagger}\Omega \mathbb{Z}_2, \hspace{1cm}  \Omega=
		\begin{pmatrix}
			0 & \delta^{\dot{\beta}}_{~\dot{\alpha}} & 0\\
			\delta_{\beta}^{~\alpha} & 0 & 0\\
			0 & 0 & \delta_j^{~i}
		\end{pmatrix}.
	\end{equation}
	
	The supertwistors $\bar{Z}=Z^\dagger \Omega$ transforming in the dual representations are given by
	\begin{equation}
		\bar{\mathbb{Z}}_A=\begin{pmatrix}
			\bar{Z}^\alpha &
			\bar{Z}_{\dot{\alpha}} &
			\bar{Z}^i\end{pmatrix}\in \mathbb{C}^{4|\mathcal{N}}
	\end{equation}
	
	The superspace is spanned by two supertwistors $\mathbb{Z}_A^{a}$ where $a\in \left\{1,2\right\}$ and its dual $\bar{\mathbb{Z}}^{\dot{a}A}$ where $\dot{a}\in$$\left\{1,2\right\}$, subject to the constraint
	\begin{equation}\label{eq:11}
		\bar{\mathbb{Z}}^{\dot{a}A}\mathbb{Z}^{a}_A=0, \hspace{1cm} a,\dot{a}\in \left\{1,2\right\}
	\end{equation}
	\paragraph{}
	The two planes in the supertwistor space suffer from the following gauge redundancies
	\begin{equation}
		\mathbb{Z}^{a}_{A}\sim \mathbb{Z}_{A}^bM^{a}_{b} \hspace{1cm} M\in GL(2,\mathbb{C})
	\end{equation} which changes the basis, and the same for the two planes in the dual supertwistor space. In order to get rid of this $GL(2,\mathbb{C})\times GL(2,\mathbb{C})$ redundancy, we can adopt gauge fixing by choosing the Poinacar\'{e} section. In that case, the supertwistors reduce to their Poincar\'{e} sections, given as follows
	\begin{equation}
		\mathbb{Z}_{A}^{a}=\begin{pmatrix}
			\delta_{\alpha}^{~a} \\
			i x_{+}^{\dot{\alpha}a}\\
			2\theta^a_{i} 
		\end{pmatrix}
		\hspace{1cm} 
		\bar{\mathbb{Z}}^{\dot{a}A}= \begin{pmatrix}
			-i x_{-}^{\dot{a}\alpha} & \delta^{\dot{a}}_{~\dot{\alpha}} & 2{\bar{\theta}}^{\dot{a}i}
		\end{pmatrix} .
	\end{equation}
	In terms of the Poincar\'{e} section the constraint (\ref{eq:11}) becomes 
	\begin{equation}
		x_{+}^{\dot{\alpha}\alpha}- x_{-}^{\dot{\alpha}\alpha}-4i\bar{\theta}^{\dot{\alpha}i}\theta^\alpha_i=0,
	\end{equation}
	where $x_\pm$ are the usual chiral and antichiral bosonic coordinates and $\theta,\bar{\theta}$ are the fermionic coordinates on the superspace. 
	
	Thus, supertwistors describe the superspace, but that involves the $GL(2,\mathbb{C})\times GL(2,\mathbb{C})$ redundancies. However, the physical quantities should be independent of this redundancy, and that motivates the introduction of bi-supertwistors. Bi-supertwistors are the physical quantities that are invariant under $SL(2,\mathbb{C})\times SL(2,\mathbb{C})$, and their gauge redundancies are fixed up to scaling. The bi-supertwistors are given by
	\begin{equation}
		\chi_{AB}\equiv \mathbb{Z}^a_A\mathbb{Z}^b_B\epsilon_{ab},\hspace{1cm} \bar{\chi}^{AB}\equiv\bar{\mathbb{Z}}^{\dot{a}A}\bar{\mathbb{Z}}^{\dot{b}B}\epsilon_{\dot{a}\dot{b}},
	\end{equation}
	which are well defined upto a rescaling 
	\begin{equation}
	(\chi , \bar{\chi}) \sim (\lambda \chi ,\bar\lambda  \bar{\chi}), \quad \lambda = \det{g}\quad \bar\lambda = \det{\bar g} .
	\end{equation}
	They satisfy a constraint in the projective null space, which is
	\begin{equation}
		\bar{\chi}^{AB}\chi_{BC}=0 
	\end{equation}
	The space described by the bi-supertwistors $\chi$ and $\bar{\chi}$, is called superembedding space.


	\subsection{Superconformal invariants}
	In superembedding space, the superconformal invariants can be expressed in terms of the supertraces of the products of $\chi$ and $\bar{\chi}$ \cite{Goldberger:2012xb,Park:1997bq,Park:1999pd}. The supertrace of bi-supertwistors is denoted as,
	\begin{equation} \label{eq:12}
		\langle \bar{2}1\rangle\equiv \bar{\chi}_2^{AB}\chi_{1BA}
	\end{equation}
	\begin{equation}
		\langle \bar{4}3\bar{2}1\rangle\equiv \bar{\chi}_4^{AB}\chi_{3BC}\bar{\chi}_2^{CD}\chi_{1DA}(-1)^{p_C}
	\end{equation}
	For example supertrace $\langle \bar{i}j\rangle$ gives the two point product
	\begin{equation}
		\langle \bar{i}j\rangle=-2x_{\bar{i}j}^2
	\end{equation}
	This can be generalized to products of more bi-supertwistors. 
	These invariants are chiral in unbarred coordinates and antichiral in barred coordinates. 
	
	The superconformal invariant ${\textbf u}$ can be constructed in superembedding space by using three points $(1,2,0)$ as follows   \cite{li2016most} ,
	\begin{equation}
		{\textbf u}=\frac{\langle \bar{1}2\rangle\langle 2\bar{0}\rangle\langle 0\bar{1}\rangle}{\langle 2\bar{1}\rangle\langle 0\bar{2}\rangle\langle 1\bar{0}\rangle}.
	\end{equation}
	Using ${\textbf u}$ we can obtain a nilpotent superconformal invariant $z$, given by, 
	\begin{equation}
		z=\frac{1-\textbf u}{1+\textbf u}=\frac{\langle \bar{1}2 \rangle\langle \bar{2}0\rangle\langle \bar{0}1\rangle-\langle \bar{2}1\rangle\langle 1\bar{0}\rangle\langle \bar{0}2\rangle }{\langle \bar{1}2 \rangle\langle \bar{2}0\rangle\langle \bar{0}1\rangle+\langle \bar{2}1\rangle\langle 1\bar{0}\rangle\langle \bar{0}2\rangle }
	\end{equation} 
	$z$ is antisymmetric under the ${\mathbb Z}_2$ that corresponds to $1\leftrightarrow 2$ interchange . For $\mathcal{N}=2$ theories, one can check that $z$ is nilpotent and $z^5$ and higher orders vanish. With the same spirit, we can construct supertraces of more points. We can obtain a new independent structure, $\langle \bar{1}2\bar{0}1\bar{2}0\rangle$, and the rest of the other structures either vanish or degenerate to 2-traces. Using these 6-traces we obtain another superconformal invariant $\textbf w'$ given by, 
	\begin{equation}
		\textbf w'=\frac{4\langle \bar{1}2\bar{0}1\bar{2}0\rangle}{(\langle \bar{1}2\rangle\langle \bar{2}1\rangle\langle\bar{1}0\rangle\langle\bar{0}1\rangle\langle\bar{0}2\rangle\langle\bar{2}0\rangle)^{\frac{1}{2}}}
	\end{equation}
	Unlike $z$, $\textbf w'$ is symmetric under $1\leftrightarrow 2$ exchange. For non-supersymmetric theories where $\theta=\bar{\theta}=0$, $\textbf w'=-1$. We define 
	\begin{equation}
		\textbf w=\textbf w'+1
	\end{equation}
	In $\mathcal{N}=1$ theories, the invariant $\textbf w$ is proportional to $z^2$ and hence it does not give any new independent superconformal invariant. But for $\mathcal{N}=2$ theories, there are two independent superconformal invariants in three point function \cite{Park:1999pd} and the same for $\mathcal{N}\geq2$.
	
	\subsection{Superconformal tensor structures}
	In superembedding space, a general superfield with spin lifts to a multi-twistor operator
	\begin{equation}
		\phi^{\dot{\beta}_1...\dot{\beta_{\bar{j}}}}_{\alpha_1...\alpha_j}\rightarrow \Phi_{B_1...B_{\bar{j}}}^{A_1...A_j}(\chi,\bar{\chi}).
	\end{equation}
	The field $\Phi$ has gauge redundancies in each index,
	\begin{equation}
		{\Phi_{B_1,..\bar{B}_{\bar{j}}}}^{A_1,..A_j}(\chi,\bar{\chi})\sim {\Phi _{B_1,..B_{\bar{j}}}}^{A_1,..A_j}(\chi,\bar{\chi})+\chi_{B_1,C}{\Lambda_{B_1,..B_{\bar{j}}}}^{CA_1..A_j}
	\end{equation}
	So it is convenient to make the field $\Phi$ index free. In order to make it index free we introduce auxiliary twistors $\mathcal{S}_A$ and $\bar{\mathcal{S}}^A$, which absorb the indices of the superembedding field. They are defined by
	\begin{equation}
		\Phi(\chi,\bar{\chi},\mathcal{S},\bar{\mathcal{S}}) \equiv \bar{\mathcal{S}}^{B_{\bar{j}}}...\bar{\mathcal{S}}^{B_{\bar{1}}} {\Phi_{B_1...B_{\bar{j}}}}^{A_1...A_j}\mathcal{S}_{A_j}...\mathcal{S}_{A_1}
	\end{equation}
	The gauge redundancy of $\Phi$ put restrictions on $\mathcal{S}$,$\bar{\mathcal{S}}$ to be transverse and null 
	\begin{equation}
		\bar{\chi}\mathcal{S}=0 \hspace{1cm} \bar{\mathcal{S}}\chi=0 \hspace{1cm} \bar{\mathcal{S}}\mathcal{S}=0
	\end{equation}
	The superfield can be recovered by
	\begin{equation}
		\phi^{\dot{\beta}_1..\dot{\beta}_{\bar{j}}}_{\alpha_1...\alpha_j}=\frac{1}{j!}\frac{1}{\bar{j}!}\left(\bar{\chi}\overrightarrow{\partial}_{\bar{\mathcal{S}}}\right)^{\dot{\beta}_1}...\left(\bar{\chi}\overrightarrow{\partial}_{\bar{\mathcal{S}}}\right)^{\dot{\beta}_{\bar{j}}}\Phi(\chi,\bar{\chi},\mathcal{S},\bar{\mathcal{S}}) \left(\overleftarrow{\partial}_{\mathcal{S}}\chi\right)_{\alpha_1}...\left(\overleftarrow{\partial}_{\mathcal{S}}\chi\right)_{\alpha_j} \mid_{\text{Poincar\'{e}}}
	\end{equation}
	The correlators are functions of tensor structures and from two points $(1,2)$ one can construct 
	\begin{equation} \label{eq:12}
		S\equiv\frac{\bar{\mathcal{S}}1\bar{2}\mathcal{S}}{\langle 1\bar{2}\rangle}, \hspace{1cm} S_{*}=S|_{1\leftrightarrow 2}\equiv\frac{\bar{\mathcal{S}}2\bar{1}\mathcal{S}}{\langle 2\bar{1}\rangle},
	\end{equation}
	where $\bar{\mathcal{S}}1\bar{2}\mathcal{S} = \bar{\mathcal{S}}^A \chi_{1AC} \bar{\chi}_2^{CD}\mathcal{S}_D$.
	
	Similarly for the points $(3,4)$ one obtains, 
	\begin{equation} \label{eq:13}
		T\equiv\frac{\bar{\mathcal{T}}3\bar{4}\mathcal{T}}{\langle 3\bar{4}\rangle}, \hspace{1cm} T_{*}=T|_{3\leftrightarrow 4}\equiv\frac{\bar{\mathcal{T}}4\bar{3}\mathcal{T}}{\langle 4 \bar{3}\rangle}.
	\end{equation}
	With (\ref{eq:12}) and (\ref{eq:13}) we construct
	\begin{equation}
		S_{\pm}=\frac{1}{2}(S \pm S_{*})
	\end{equation}
	and 
	\begin{equation}
		T_{\pm}=\frac{1}{2}(T \pm T_{*})
	\end{equation}
	$S_{+}$ is nilpotent and vanishes for $\theta_{i}=0$. For $\mathcal{N}=1$ superembedding space, $S_{+}^2$ becomes $0$ and for $\mathcal{N}=2$ superembedding space, $S_{+}^3$ vanishes, so the tensor structures terminate at $S_{+}^2$. For the points (1,2) we have the tensor structures,
	\begin{equation} \label{eq:14}
		S_{-}^l\sim\frac{1}{2}\left(S^l+(-1)^lS_{*}^l\right),
	\end{equation}
	\begin{equation}
		S_{+}S_{-}^{l-1}=\frac{1}{2l}\left(S^{l}-(-1)^lS_{*}^l\right),
	\end{equation}
	\begin{equation}
		S_{+}^2S_{-}^{l-2}=\frac{1}{4(l-1)}\left(S^l+S^{l-1}S_{*}+(-1)^lSS_{*}^{l-1}+(-1)^lS_{*}^l\right)
	\end{equation}
	and for the tensor structures associated with the points $(3,4)$ are:
	\begin{equation}  \label{eq:15}
		T_{-}^l\sim\frac{1}{2}\left(T^l+(-1)^lT_{*}^l\right),
	\end{equation}
	\begin{equation}\label{eq:16}
		T_{+}T_{-}^{l-1}=\frac{1}{2l}\left(T^{l}-(-1)^lT_{*}^l\right),
	\end{equation}
	The parity of the tensor structure has been taken care of by $(-1)^l$ under the $1\leftrightarrow 2$ exchange. $S_{-}^l$ and $S_{+}^2S_{-}^{l-2}$ have same parity while the parity of $S_{-}^l$ and $S_{+}S_{-}^{l-1}$ are completely opposite to each other under the exchange $1\leftrightarrow 2$. Similarly, under the coordinate exchange of $3\leftrightarrow 4$, $T_{-}^l$ and  $T_{+}T_{-}^{l-1}$ have opposite parities, as follows from (\ref{eq:15}) and (\ref{eq:16}). It may be noted that (\ref{chiral3point}) involves $T^l$, which can be expressed in terms of (\ref{eq:15}) and (\ref{eq:16}) only.  
	
	\subsection{Supershadow approach} 
	We follow \cite{fitzpatrick2014covariant,khandker2014n} for supershadow formalism. For the nonsupersymmetric case the shadow formalism was introduced in \cite{simmons2014projectors}. The main idea behind this approach is that given an operator $\mathcal{O}:(\frac{l}{2},\frac{\bar{l}}{2},q,\bar{q})$,where $q$ and $\bar{q}$ are the superconformal weights of the operator,
	\begin{equation}
		q\equiv\frac{1}{2}\left(\Delta+\frac{3}{2}r\right), \hspace{1cm }\bar{q}\equiv \frac{1}{2}\left(\Delta-\frac{3}{2}r\right),
	\end{equation}  one can construct another operator $\tilde{\mathcal{O}}:(\frac{\bar{l}}{2},\frac{l}{2}, 2 - \mathcal{N}-q, 2-\mathcal{N} - \bar{q})$ which has the same eigenvalue of the superconformal Casimir as that of $\mathcal{O}:(\frac{l}{2},\frac{\bar{l}}{2},q,\bar{q})$. The operator $\tilde{\mathcal{O}}$ is called the shadow operator of the operator $\mathcal{O}$. 
	
Consider a given multiplet $\mathcal{O}$ of superconformal weights $(q,\bar{q})$ of spin $(j, \bar{j})$, the shadow operator \cite{fitzpatrick2014covariant} is given by
	\begin{equation}\label{eq:3}
		\tilde{\mathcal{O}}(1,\bar{1},\mathcal{S},\mathcal{\bar{S}})=\int D[2,\bar{2}]\frac{\mathcal{O}^\dagger(2,\bar{2},2\bar{\mathcal{S}},\bar{2}\mathcal{S})}{\langle 1\bar{2}\rangle^{2-\mathcal{N} - q + j }\langle \bar{1}2\rangle^{2-\mathcal{N}-\bar{q}+\bar{j}}}.
	\end{equation}
	In the above equation, the integration measure is in superembedding space, and the operator ${\mathcal{O}^\dagger}$ of spin $( \bar{j} , j )$ is the  Lorentz conjugate of the operator $\mathcal{O}$. Since the operators $\tilde{\mathcal{O}}$ and $\mathcal{O}^{\dagger}$ are related through a non-local, linear transformation, they share the same Lorentz indices.

	Considering that a general superfield $\varphi$ in $\mathcal{N}=2$ superembedding space of quantum numbers $(\Delta,j , \bar{j} , R=0, r )$ satisfies the homogeneity given by,
	\begin{equation}
		\varphi(\lambda\chi,\bar{\lambda}\bar{\chi})=\lambda^{-q-j}\bar{\lambda}^{-\bar{q}- \bar{j}}\varphi(\chi,\bar{\chi}).
	\end{equation}
	 One can check the homogeneity of the shadow operator.
	
	The homogeneity of the shadow operator, as per (\ref{eq:3}) is given by\cite{li2020superconformal} 
	\begin{equation}\label{eq:4}
		\tilde{\mathcal{O}}(\lambda\chi,\bar{\lambda}\bar{\chi},a\mathcal{S},\bar{a} \bar{\mathcal{S}})=\lambda^{-(2-\mathcal{N} - q + j )}\bar{\lambda}^{-( 2 - \mathcal{N} - \bar{q} +\bar{ j})} a^{2j}\bar{a}^{2\bar{j}}\tilde{\mathcal{O}}(\chi,\bar{\chi},\mathcal{S},\bar{\mathcal{S}}).
	\end{equation}
	Thus it transforms as a superconformal multiplet with quantum numbers $(j , \bar{j}, 2-\mathcal{N}-q,2-\mathcal{N}-\bar{q})$.
	
	Given an operator $\mathcal{O}$ and the corresponding shadow operator $\tilde{\mathcal{O}}$ one can construct a conformal projector 
	\begin{equation}\label{eq:5}
		\big|\mathcal{O}\big|=\frac{1}{(2j)!^2(2\bar{j})!^2}\int D[0,\bar{0}] \ket{\mathcal{O}(0,\bar{0},\mathcal{S},\bar{\mathcal{S}})}\left(\overleftarrow{\partial_{\mathcal{S}}}0\overrightarrow{\partial_{\mathcal{T}}}\right)\left(\overleftarrow{\partial_{\bar{\mathcal{S}}}}0\overrightarrow{\partial_{\bar{\mathcal{T}}}}\right)\bra{\tilde{\mathcal{O}}(0,\bar{0},\mathcal{T},\bar{\mathcal{T}})} \Big|_M
	\end{equation}
	In the above equation, $M$ refers to monodromy projection \cite{simmons2014projectors}. 
	In general, the expression would involve a contribution from the shadow operators. As explained in \cite{simmons2014projectors}, by restricting to specific monodromy projection, one can restrict the region of integration and avoid erroneous shadow contributions.
	Given a four-point correlator, this projector can be used to project the correlation function into the superconformal partial waves corresponding to operator $\mathcal{O}$. In particular, considering the present correlator $\langle \mathcal{J J}\Phi\Phi^\dagger\rangle$ can be projected to the partial waves,
	\begin{equation}\label{eq:6}
		\mathcal{W}_{\mathcal{O}}\propto \langle\mathcal{J}\mathcal{J}|\mathcal{O}|\Phi\Phi^\dagger\rangle \sim \int D[0,\bar{0}]\langle\mathcal{J}\mathcal{J}\mathcal{O}\rangle\overleftrightarrow{\mathcal{D}}_{j , \bar{j}}  \langle\tilde{\mathcal{O}}\Phi\Phi^\dagger\rangle
	\end{equation}
	where,
	\begin{equation}
		\overleftrightarrow{\mathcal{D}}_{j , \bar{j} }=\frac{1}{(2j)!^2(2\bar{j})!^2}(\partial_{\mathcal{S}}0\partial_{\mathcal{T}})^{ 2 j }(\partial_{\bar{\mathcal{S}}}\bar0\partial_{\bar{\mathcal{T}}})^{ 2 \bar{j} }.
	\end{equation}
	We will compute the partial waves using this projector. 
	\section{Three point function and OPE coefficient}
	
	Our objective is to obtain an expression for a mixed four-point correlator $\langle J(x_1)J(x_2)\phi(x_3)\phi^\dagger(x_4)\rangle $ using the techniques of shadow formalism. The procedure, as we have already explained, involves two kinds of three point functions. The first one consists of two scalar operators of stress tensor multiplet with a third arbitrary operator, whereas the second kind consists of chiral-antichiral operators with a third arbitrary operator. In this section, we will briefly sketch how to determine these three point functions.
	
	From the OPE selection rule, we can get the possible third operator $\mathcal{O}$ appearing in the three point functions which will give a non-zero expression. Considering the OPE for both $J$ and the chiral operator, one can observe that the number of possible multiplets in this mixed case is quite limited, compared to the case of four point correlators of $J$. It consists of the long multiplets $\mathcal{A}_{0,0(\frac{l}{2},\frac{l}{2})}^{\Delta}$. The $\mathcal{C}$ type multiplets either disappear or contain a unique conformal block whose solutions are closely related to long multiplets. We have commented on this issue at the end of section 5. Therefore, in this present work, we are interested in the long multiplets.
	\paragraph{}
	
	The procedures to obtain the three point functions $\langle\mathcal{J}\mathcal{J}\mathcal{O}\rangle$ and $\langle\Phi\Phi^\dagger\mathcal{O}\rangle$ have been discussed in \cite{liendo2016stress}. It involves writing down the most general ansatz consistent with the superconformal symmetry and the relevant equation. Then, imposing the reflection symmetry, $z_1 \leftrightarrow z_2$, one can determine the coefficients. A similar procedure is adopted in superembedding space \cite{fitzpatrick2014covariant,li2020superconformal}, which consists of the following two steps :
	
	$\bullet$ Write the general ansatz using homogeneity and reflection symmetry.
	
	$\bullet$ Solve the coefficients by imposing the relevant equations.
		
	The fundamental constituents of a three point function in superembedding space are the superconformal invariants $z$, $w$ and the tensor structures $S_{-}^l$, $S_{+}S_{-}^{l-1}$, $S_{+}^2S_{-}^{l-2}$.  In what follows, we will discuss the three point functions in terms of these tensor structures.

	\subsection{Three point function $\langle \mathcal{J}\mathcal{J}\mathcal{O}\rangle$ for long multiplet $\mathcal{A}_{0,0(\frac{l}{2},\frac{l}{2})}^{\Delta}$}
	The three point function $\langle \mathcal{J}\mathcal{J}\mathcal{O}\rangle$ in terms of the tensor structures has been discussed in \cite{li2020superconformal} for different multiplets. For our case, only the long multiplet is relevant and we review the computation for the exchange of long multiplet. These are determined using superconformal symmetry, reflection symmetry, and the relevant equation \cite{li2020superconformal}.  The three point correlations $\langle {\mathcal {J J} O}\rangle$ involves two linearly independent superconformal invariants \cite{Park:1999pd} and new tensor structures \cite{li2020superconformal}. 
It turns out that the three point functions are different for odd/even values of spin due to reflection symmetry. 
\paragraph{}

	For odd spin long multiplet the most general expression for the three point function, in terms of the tensor structures, consistent with the reflection symmetry is given by 
	\begin{equation} \label{eq:1}
		\langle\mathcal{J}(1,\bar{1})\mathcal{J}(2,\bar{2})\mathcal{O}(0,\bar{0})\rangle=\frac{S^l_{-}(\lambda_1 z+\lambda_3 z^3)+S_{+}S_{-}^{l-1}(\lambda_0+\lambda_2 z^2)}{(\langle\bar{1}2\rangle\langle\bar{2}1\rangle)^{1-\frac{1}{4}(\Delta+l)}(\langle\bar{0}1\rangle\langle\bar{1}0\rangle\langle\bar{0}2\rangle\langle\bar{2}0\rangle)^{\frac{1}{4}(\Delta+l)}}. 
	\end{equation}
	The parity of $S_{-}^l$ and $z$ are odd under $1\leftrightarrow 2$ exchange.
	There can be some extra contribution of the tensor structure $S_{+}^2S_{-}^{l-2}z$. However, they will not give any new independent term, and the other superconformal invariant $\textbf w$ does not appear in the above equation. 
	
	Application of the conservation equation will impose additional  restrictions on the three point function ansatz and the OPE coefficients $\lambda_i$ are fixed to the following \cite{li2020superconformal}, 
	\begin{equation}
		\vec\lambda_{\mathcal{O}}=\lambda_{\mathcal{O}}\left(1,-\frac{\Delta+l}{2(\Delta-2)},\frac{(\Delta+l)^2(2l^2-2l-8+6\Delta+l\Delta-\Delta^2)}{8(\Delta-2)},-\frac{(\Delta+l)(5l^2-8+16\Delta+2l\Delta-3\Delta^2)}{48(\Delta-2)}\right)
	\end{equation}
	Expanding the equation (\ref{eq:1}) with the OPE coefficients above it will be consistent with the result of \cite{liendo2016stress}. The coefficients are multiplied by an overall factor $(4+l-\Delta)$ compared with the expressions in \cite{liendo2016stress} to avoid the unphysical poles in the above coefficients. There is another pole at $\Delta=2$ but this is below the unitarity bounds of any multiplet of odd spin.
	\paragraph{}
	
	In the case of even $l$, the ansatz for the three point function from supersymmetry and reflection symmetry is
	\begin{equation}\label{eq:8}
		\langle\mathcal{J}(1,\bar{1})\mathcal{J}(2,\bar{2})\mathcal{O}(0,\bar{0})\rangle=\frac{S^l_{-}(\lambda_0+\lambda_2 z^2+\lambda_3 \textbf w+\lambda_4 z^4)+S_{+}S_{-}^{l-1}\lambda_1 z+S^2_{+}S_{-}^{l-2}\lambda_5}{(\langle\bar{1}2\rangle\langle\bar{2}1\rangle)^{1-\frac{1}{4}(\Delta+l)}(\langle\bar{0}1\rangle\langle\bar{1}0\rangle\langle\bar{0}2\rangle\langle\bar{2}0\rangle)^{\frac{1}{4}(\Delta+l)}}.
	\end{equation}
	Unlike the case of odd $l$, the coefficients $\lambda_i$s, in this case, cannot be fixed up to an overall term due to the presence of two independent superconformal invariants $z$ and $\textbf w$. Rather, the conservation equation leads to two independent solutions, which are given by \cite{li2020superconformal},
	\begin{dmath}
		\vec\lambda^{(1)}_{\mathcal{O}}=\lambda^{(1)}_{\mathcal{O}}\left(1,\frac{(\Delta+l)(2-\Delta)}{2},\frac{(\Delta+l)^2}{8},0,\frac{((\Delta+l)(64-3l^3+(\Delta-4)^2)\Delta-l^2(5\Delta-8)-l(\Delta^2-48))}{384},\\
		\frac{(2+l-\Delta)(2-l-\Delta)}{2} \right),
	\end{dmath}
	and 
	\begin{multline}
		\vec{\lambda}^{(2)}_{\mathcal{O}}=\lambda_{\mathcal{O}}^{(2)}(0,\frac{(\Delta+l)(6+l-3\Delta)}{2},\frac{(\Delta+l)(l+3\Delta)}{8},-\frac{(\Delta+l)(4+l-\Delta)}{4},\\
		\frac{(\Delta+l)(3(4+3\Delta)-l(1-2\Delta+2l(3+l+\Delta)))}{96},\frac{(3(\Delta-2)^2-2l-l^2)}{2} ).
	\end{multline}
	These two solutions once again are rescaled by multiplying a factor of $(4+l-\Delta)$ to avoid unphysical poles in the coefficients.
	
	The three point function $\langle\mathcal{J}\mathcal{J}\mathcal{J}\rangle$ has been studied in superspace \cite{kuzenko2000correlation}. They have determined the restrictions on the general structure of the three point function imposed by the $\mathcal{N}=2$ superconformal symmetry, which admits two independent structures. In Section 5.2 we see that the conformal blocks will appear with two different sets of $\lambda^{(i)}$s.
	
	\subsection{The three point function $\langle \Phi\Phi^{\dagger}\mathcal{O}\rangle$ for long multiplet $\mathcal{A}_{0,0,(\frac{l}{2},\frac{l}{2})}^{\Delta}$}
	The chiral-antichiral three point function is given in \cite{fitzpatrick2014covariant}.
	\begin{equation}\label{chiral3point}
		\langle\Phi(3,\bar{3})\Phi^{\dagger}(4,\bar{4})\mathcal{O}(\chi_0,\chi_{\bar{0}},\mathcal{T},\bar{\mathcal{T}}) \rangle
		\propto\frac{(\bar{\mathcal{T}}3\bar{4}\mathcal{T})^l}{\langle{3\bar{4}}\rangle^{q_{\phi}+\frac{1}{4}(\Delta-l)}\langle{4\bar{3}}\rangle^{{\frac{1}{4}(\Delta-l)}}(\langle\bar{0}3\rangle\langle\bar{3}0\rangle\langle\bar{0}4\rangle\langle\bar{4}0\rangle)^{\frac{1}{4}(\Delta-l)}}
	\end{equation}
	For the purpose of computation of the four point correlator, we require the three point function with the shadow operator $\langle\tilde{\mathcal{O}}\Phi{\Phi}^{\dagger}\rangle$, which can be obtained by setting $\Delta\rightarrow-\Delta$.
	
	As explained in \cite{khandker2014n}, in order to express the above expression in terms of the tensor structures and superconformal invariants mentioned above, we need to use the identity
	\begin{equation}
		\left(\frac{\langle 3\bar{4}\rangle}{\langle 3\bar{0}\rangle\langle 0\bar{4}\rangle}\right)^{2\delta}=\left(\frac{\langle 3\bar{4}\rangle\langle 4\bar{3}\rangle}{\langle 3\bar{0}\rangle\langle 0\bar{4}\langle 4\bar{0}\rangle\langle 0\bar{3}\rangle}\right)^{\delta}(1-2\delta\tilde{z}+2\delta^2{\tilde{z}}^2-\frac{2\delta}{3}(2\delta^2+1){\tilde{z}}^3+\frac{2\delta^2}{3}(\delta^2+2){\tilde{z}}^4).
	\end{equation}
	We are truncating after order 4 of the superconformal invariant $z$, since the terms of higher order in $z$ vanish .
	Hence the most general ansatz we obtained for spin $l$
	\begin{equation}\label{eq:9}
		\begin{split} 
			\langle\Phi(3,\bar{3})\Phi^{\dagger}(4,\bar{4})\mathcal{O}(0,\bar{0})\rangle & =
			\frac{( T^l_{-} + l ~T_{+}T_{-}^{l-1})( \tilde{\lambda}_0+\tilde{\lambda}_1 \tilde{z} + \tilde{\lambda}_2 \tilde{z}^2 + \tilde{\lambda}_3 \tilde{z}^3 + \tilde{\lambda}_4 \tilde{z}^4) 
			}{\langle 3\bar{4}\rangle^{\frac{1}{4}(\Delta -l)}\langle 4\bar{3}\rangle^{\frac{1}{4}(\Delta-l)}(\langle\bar{0}3\rangle\langle\bar{3}0\rangle\langle\bar{0}4\rangle\langle\bar{4}0\rangle)^{\frac{1}{4}(\Delta-l)}}.
		\end{split}
	\end{equation}	
	The above three point function is consistent with superconformal symmetry and reflection symmetry. The superconformal blocks 	$\mathcal{G}_{\Delta,l}^{\mathcal{N}=2|JJ;\phi\phi^\dagger}$ we obtained using the above three point function, follows the $\mathcal{N}=1$ decompositions as mentioned in Section 6.
	The OPE coefficients  are
	\begin{equation}\begin{split}
			\vec{\tilde\lambda}_{\phi{\phi^\dagger}\mathcal{O}} &={\tilde\lambda}_{\mathcal{O}}\left(1,-\frac{(\Delta+l)}{2},\frac{(\Delta+l)^2}{8},-\frac{(\Delta+l)}{6}(1+\frac{(\Delta+l)^2}{8}),\frac{(\Delta+l)^2}{24}(2+\frac{(\Delta+l)^2}{16}) \right).
	\end{split}\end{equation}
	For shadow three point function $\langle\tilde{\mathcal{O}}\Phi{\Phi}^{\dagger}\rangle$ the OPE coefficients will be rewritten by replacing $\Delta$ to $-\Delta$.
	
	%
	\section{Computation of superconformal partial waves}
	In this section we calculate superconformal partial waves $ \mathcal{W}_{\mathcal{O}} $ for the mixed correlation function \\
	$\langle J(1,\bar{1})J(2,\bar{2})\phi(3,\bar{3})\phi^{\dagger}(4,\bar{4}) \rangle$, where $J$, $\phi$ and $\phi^{\dagger}$ are the external operator representing superconformal primary operators of stress-tensor, chiral and anti-chiral multiplets respectively and are given by
	\begin{equation*}
		J(x)=\mathcal{J}(x,\theta,\bar{\theta})\bigg|_{\theta_i=\bar{\theta}^i=0}\quad\text{and}\quad 
		\phi(x)=\Phi(x,\theta,\bar{\theta})\bigg|_{\theta_i=\bar{\theta}^i=0}.
	\end{equation*} and similarly for $\phi^\dagger$. 
	
	We use the method of projectors where the  superconformal partial wave $\mathcal{W}_{\mathcal{O}}$ under the exchange of a real superfield $\mathcal{O}$ with quantum numbers $(\Delta, \frac{l}{2},\frac{l}{2},R=0,r=0)$ ( or equivalently $(\frac{l}{2},\frac{l}{2},\frac{\Delta}{2},\frac{\Delta}{2})$ in $(j, \bar{j} , q, \bar{q})$ notation ) for the $s$-channel or the direct channel is given by
	\begin{equation}\label{eq:6}
		\mathcal{W}_{\mathcal{O}}\propto \langle\mathcal{J}\mathcal{J}|\mathcal{O}(0,\bar{0})| \Phi \Phi^\dagger \rangle \sim \int D[0,\bar{0}]\langle\mathcal{J}\mathcal{J}\mathcal{O}(0,\bar{0},\mathcal{S},\bar{\mathcal{S}})\rangle\overleftrightarrow{\mathcal{D}}_{j , \bar{j} }\langle\tilde{\mathcal{O}}(0,\bar{0},\mathcal{S},\bar{\mathcal{S}})\Phi \Phi^\dagger\rangle
	\end{equation}
	where,
	\begin{equation}
		\overleftrightarrow{\mathcal{D}}_{j , \bar{j}}=\frac{1}{(2j_1)!^2(2j_2)!^2}(\partial_{\mathcal{S}}0\partial_{\mathcal{T}})^{2j }(\partial_{\bar{\mathcal{S}}}\bar0\partial_{\bar{\mathcal{T}}})^{2\bar{j}}.
	\end{equation}
	The superintegrand consists of two three-point functions $\langle\mathcal{J}\mathcal{J}\mathcal{O}\rangle$ and the supershadow of  $\langle\Phi\Phi^{\dagger}\mathcal{O}\rangle$. 
	The superconformal integral was presented in \cite{fitzpatrick2014covariant}.
	\subsection{Result for $\mathcal{A}^\Delta_{0,0(\frac{l}{2},\frac{l}{2})}$ with $l$ odd}
	We begin with the long multiplet with odd $l$ as the exchange operator,  and as follows from  (\ref{eq:6}) the superconformal partial wave is,
	\begin{equation}\begin{split}
			{\mathcal W}_{\mathcal O} &\propto \langle {\mathcal J} (1,\bar{1}) {\mathcal J} (2,\bar{2}) |{\mathcal O}| \Phi(3,\bar{3}) \Phi^\dagger (4,\bar{4})\rangle
			\\
			&\propto \int D[0,\bar{0}] \langle {\mathcal J}(1,\bar{1})\mathcal{J}(2,\bar{2})\mathcal{O}(0,\bar{0},\mathcal{S},\mathcal{\bar{S}})\rangle\overleftrightarrow{\mathcal{D}}_l\langle \tilde{\mathcal O}(0,\bar{0},{\mathcal T},\tilde{\mathcal T})\Phi(3,\bar{3})\Phi^\dagger (4,\bar{4})\rangle.
		\end{split}
	\end{equation}
	The three point functions in terms of the tensor structures have already been given in (\ref{eq:1})) and (\ref{eq:9}). Substituting these expressions in the equation above, the superconformal partial wave (for odd $l$ ) can be written as
	\begin{equation}\label{17}
		\mathcal{W}_{\mathcal{O}}\propto \frac{1}{(\langle 1\bar{2}\rangle\langle2\bar{1}\rangle)^{1-\frac{1}{4}(\Delta+l)}(\langle 3\bar{4}\rangle\langle 4\bar{3}\rangle)^{1-\frac{1}{4}(l-\Delta)}}\times
		\\ \int D[0,\bar{0}]\frac{\mathcal{N}^{full}_l}{(\langle 1\bar{0}\rangle\langle 2\bar{0}\rangle\langle0\bar{1}\rangle\langle0\bar{2}\rangle)^{\frac{1}{4}(l+\Delta)}(\langle 3\bar{0}\rangle\langle 4\bar{0}\rangle\langle 0\bar{3}\rangle\langle0\bar{4}\rangle)^{\frac{1}{4}(l-\Delta)}}
	\end{equation}
	where 
	\begin{equation}
		\mathcal{N}_l^{Full}=\mathcal{N}_l^{(1)}+\mathcal{N}_l^{(2)}+\mathcal{N}_l^{(3)}+\mathcal{N}_l^{(4)}
	\end{equation}
	Here,\begin{equation}
		\begin{split} \label{18}
			\mathcal{N}_l^{(1)} &= ({\lambda_1}\tilde{\lambda}_1 z\tilde{z}+{\lambda_1}\tilde{\lambda}_3 z\tilde{z}^3+\lambda_3\tilde{\lambda}_1z^3 \tilde{z}) S_{-}^l\overleftrightarrow{\mathcal{D}_l}T_{-}^l \\
			\mathcal{N}_l^{(2)} &= l  ({\lambda_1}\tilde{\lambda}_0 z+{\lambda}_1\tilde{\lambda}_2 z\tilde{z}^2+ {\lambda}_3\tilde{\lambda}_0 z^3)S_{-}^l \overleftrightarrow{\mathcal{D}_l}T_{+}T_{-}^{l-1}\\
			\mathcal{N}_l^{(3)} &= ({\lambda_0}\tilde{\lambda}_1 \tilde{z} + {\lambda}_0 \tilde{\lambda}_3 \tilde{z}^3+\lambda_2\tilde{\lambda}_1 z^2\tilde{z}) S_{+}S_{-}^{l-1} \overleftrightarrow{\mathcal{D}_l}T_{-}^l \\
			\mathcal{N}_l^{(4)} &= l ({\lambda_0}\tilde{\lambda}_0 + {\lambda_0}\tilde{\lambda}_2 \tilde{z}^2+\lambda_0\tilde{\lambda}_4 \tilde{z}^4 + {\lambda}_2 \bar{\lambda}_0 z^2+{\lambda}_2\bar{\lambda}_2 z^2 \tilde{z}^2) S_{+}S_{-}^{l-1} \overleftrightarrow{\mathcal{D}_l}T_{+}T_{-}^{l-1} \\
	\end{split}\end{equation}
	and the tensor structures are given by \cite{li2020superconformal}, 
	\begin{equation}
		S_{-}^l \overleftrightarrow{\mathcal{D}_l}T_{-}^l=\left(\frac{\mathcal{N}_l}{4\langle 12\rangle^l\langle 34\rangle^l}+(-1)^l(1\leftrightarrow 2)+(-1)^l(3 \leftrightarrow 4)\right)
	\end{equation}
	\begin{equation}
		S_{-}^l \overleftrightarrow{\mathcal{D}_l}T_{+}T_{-}^{l-1}=\left(\frac{\mathcal{N}_l}{4 l \langle 12\rangle^l\langle 34\rangle^l}+(-1)^l(1\leftrightarrow 2)-(-1)^l(3 \leftrightarrow 4)\right)
	\end{equation}
	\begin{equation}
		S_{+}S_{-}^{l-1} \overleftrightarrow{\mathcal{D}_l}T_{-}^l=\left(\frac{\mathcal{N}_l}{4 l \langle 12\rangle^l\langle 34\rangle^l}-(-1)^l(1\leftrightarrow 2)+(-1)^l(3 \leftrightarrow 4)\right)
	\end{equation}
	\begin{equation}
		S_{+}S_{-}^{l-1} \overleftrightarrow{\mathcal{D}_l}T_{+}T_{-}^{l-1}=\left(\frac{\mathcal{N}_l}{4 l^2 \langle 12\rangle^l\langle 34\rangle^l}-(-1)^l(1\leftrightarrow 2)-(-1)^l(3 \leftrightarrow 4)\right)
	\end{equation}
	In the above equations, the tensor structures are expanded in terms of $\mathcal{N}_l$, where
	\begin{equation*}
		\mathcal{N}_l=\frac{1}{(l!)^2}(\partial_{\mathcal{S}}0\partial_{\mathcal{T}})^l(\mathcal{S}\bar{2}1\bar{0}3\bar{4}\mathcal{T})^l = (-1)^l s^{\frac{l}{2}}C^{(1)}_l(t)
	\end{equation*}

	in which 
	\begin{equation*}
		t\equiv\frac{\langle\bar{2}1\bar{0}3\bar{4}0\rangle}{2\sqrt{y}} \hspace{1cm} s\equiv\frac{1}{2^{16}}\langle\bar{0}1\rangle\langle\bar{2}0\rangle\langle\bar{0}3\rangle\langle\bar{4}0\rangle\langle\bar{2}1\rangle\langle\bar{4}3\rangle.
	\end{equation*}
	We are interested in the case where the superfields in the four point function are restricted to their lowest component fields. We refer to these superfields as external fields in contrast with the exchanged operator ${\mathcal O}$.
	Setting the fields to their lowest component amounts to setting their grassmann coordinates to zero, {\it i.e.} $\theta_{\text{ext}}=\bar\theta_{\text{ext}}=0$,
	$\mathcal{S}$ becomes $S$ and $\mathcal{T}$ becomes $T$ and the bi-supertwistors $\chi_{\alpha\beta}$ and $\bar{\chi}^{\alpha\beta}$ becomes bi-twistors $X_{\alpha\beta}$ and $\bar{X}^{\alpha\beta}$, which are $4\times 4$ anti-symmetric matrices with twistor indices $\alpha,\beta$ \cite{khandker2014n,li2020superconformal}. The bi-twistors $X_{\alpha\beta}$ and $\bar{X}^{\alpha\beta}$ can be defined in terms of the vectors $X_m$ of the conformal group as follows
	\begin{equation*}
		X^{\alpha\beta}=\frac{1}{2}X_m\Gamma^{m\alpha\beta} \hspace{1cm} 	\bar{X}_{\alpha\beta}=\frac{1}{2}X^m\tilde{\Gamma}_{m\alpha\beta}
	\end{equation*} and the supertraces $\langle i\bar{j}\rangle$ reduces to the inner products of the 6D vectors
	$X_{i}$ and $X_{j}$, {\it i.e.},
	\begin{equation*}
		\langle i\bar{j}\rangle=-X_i.X_j=	\frac{1}{2} X_{ij}.
	\end{equation*} After setting the external $\theta$s to zero the tensor structure becomes
	\begin{equation*}
		\mathcal{N}_l\equiv N_l=(-1)^l s_{0}^{\frac{l}{2}}C_{l}^{(1)}(t_0)
	\end{equation*}
	where,
	\begin{equation*}
		s\rightarrow s_0=\frac{1}{2^{12}}X_{10}X_{20}X_{30}X_{40}X_{12}X_{34}
	\end{equation*}
	\begin{equation*}
		t\rightarrow t_0=-\frac{X_{13}X_{20}X_{40}}{2\sqrt{X_{10}X_{20}X_{30}X_{40}X_{12}X_{34}}}-(-1)^l(1\leftrightarrow{2})-(-1)^l(3\leftrightarrow{4}).
	\end{equation*}
	In eq.(\ref{17}), it is convenient to integrate over the fermionic components of $(0,\bar{0})$ at first \cite{li2020superconformal}, which leads to an integration in embedding space and the superconformal integration becomes \begin{equation}
		\mathcal{W}_{\mathcal{O}}|_{\theta_{\text{ext}}=0} \propto\frac{1}{X_{12}^{2-\frac{1}{2}(\Delta+l)}X_{34}^{2-\frac{1}{2}(\l-\Delta)}}\int D^4X_0\partial_{\bar{0}}^4\frac{\mathcal{N}_l}{D_l}|_{\bar{0}=0}
	\end{equation}
	where $\frac{1}{D_l}$ is given by,
	\begin{equation}
		\frac{1}{D_l}\equiv \frac{1}{(X_{1\bar{0}}X_{0\bar{1}}X_{2\bar{0}}X_{0\bar{2}})^{\frac{1}{4}(l+\Delta)}(X_{3\bar{0}}X_{0\bar{3}}X_{4\bar{0}}X_{0\bar{4}})^{\frac{1}{4}(l-\Delta)}}.
	\end{equation}
	\paragraph{}
	
	$\mathcal{N}^{Full}_l$ contains several terms with $z$ and $\bar{z}$. One can carry out the derivatives ($\partial_{\bar{0}}^4$) in a straightforward manner, and the necessary formulas are collected in Appendix A, (\ref{A1}-\ref{A12}). However, the derivatives of the terms corresponding to that in (\ref{18}) lead to long expressions for each of the terms. To avoid clutter, we are not including all the expressions. As one example, we are giving the result for the following term,
	\begin{center}
		$\lambda_1\tilde\lambda_1 z\tilde{z}\frac{1}{{D}_l}S_{-}^l \overleftrightarrow{\mathcal{D}_{l}}T_{-}^l$.
	\end{center}
	Integrating out the fermionic variables we obtain the conformal integration in embedding space:
	\begin{dmath}
		(\partial^2_{\bar{0}} \partial^2_{\bar{0}} z\tilde z)\frac{1}{D_l}N_l+(\partial^n_{\bar{0}} \partial^2_{\bar{0}} z\tilde z)(\partial_{\bar{0}n}\frac{1}{D_l})N_l+2(\partial^2_{\bar{0}} z\tilde{z})({\partial}^2_{\bar{0}}\frac{1}{D_l})N_l+4(\partial^{n}\partial^{m} z\tilde{z})(\partial_{n}\partial_{m}\frac{1}{D_l})N_l+
		4(\partial^n_{\bar{0}} \partial^2_{\bar{0}} z\tilde z)\frac{1}{D_l}(\partial_{\bar{0}n}N_l)+8(\partial^{n}\partial^{m} z\tilde{z})(\partial_{n}\frac{1}{D_l}\partial_{m}N_l)+4(\partial^2 z\tilde{z})(\partial_{n}\frac{1}{D_l}\partial_{m}N_l)+4(\partial^{n}\partial^{m} z\tilde{z})\frac{1}{D_l}\partial_{n}\partial_{m}N_l+(-1)^l(1\leftrightarrow{2})+(-1)^l(3\leftrightarrow{4}).
	\end{dmath}
	After carrying out the partial derivatives, we set $\theta_{ext}=0$ and the above expression becomes,
	\begin{dmath}
		2\bigg(\Omega_{-}\Big(\big((\large \Delta + l \large)^2+8\big)\frac{X_{12}}{X_{1\bar{0}}X_{2\bar{0}}}+\big(( \Delta - l )^2+8\big)\frac{X_{34}}{X_{3\bar{0}}X_{4\bar{0}}}\Big)\\+(2+l-\Delta)(2+l+\Delta)(\Omega_{A}\Omega_{B}+\Omega_{-}\Omega_{+})\bigg)
		+\frac{\Delta +l+2}{2}\frac{l}{D_l}\frac{X_{12}}{X_{1\bar{0}}X_{2\bar{0}}}(\Omega_{B}P1-\Omega_{-}P3)-\frac{\Delta -l-2}{2}\frac{l}{D_l}\frac{X_{34}}{X_{3\bar{0}}X_{4\bar{0}}}(\Omega_{A}R1-\Omega_{-}R3)+\frac{l(l-1)}{8}\frac{1}{D_l}\frac{X_{12}}{X_{1\bar{0}}X_{2\bar{0}}}\frac{X_{34}}{X_{3\bar{0}}X_{4\bar{0}}}(P0R3+P1R1+P2R2+P3R0).
	\end{dmath}
	where the expressions for $\Omega_\pm$, $\Omega_A$, $\Omega_B$, $P0, P1, P2, P3$ and $R0, R1, R2, R3$ are presented in Appendics B.
	
	The next step is to carry out the conformal integrations. Once again it is straightforward, and the necessary formulas are collected in Appendix B. Suppressing the kinematic factors, the partial wave of the four point function
	$\langle J(1,\bar{1})J(2,\bar{2})\phi(3,\bar{3})\phi^{\dagger}(4,\bar{4})\rangle$ for general long multiplet $\mathcal{A}_{0,0(\frac{l}{2},\frac{l}{2})}^\Delta$ with odd spin $l$ is expressed as given below,
	\begin{dmath} \label{eq:p}
		\mathcal{G}_{\Delta,l,odd}^{\mathcal{N}=2|JJ;\phi\phi^{\dagger}}=\lambda_{\mathcal{O}}\tilde{\lambda}_{\tilde{\mathcal O}}\bigg[-\frac{1}{2(\Delta+l+1)(\Delta+l+3)}g_{\Delta+3,l+1}\\+\frac{8(\Delta+1)(\Delta-l)}{(\Delta-1)(\Delta-l-2)(\Delta+l)(\Delta+l+2)}g_{\Delta+1,l+1}
		-\frac{8(\Delta+1)(l+2)}{l(\Delta-1)(\Delta+l)^2}g_{\Delta+1,l-1}
		+\frac{(\Delta-l)(\Delta-l-2)(l+2)}{2l(\Delta-l+1)(\Delta+l)(\Delta+l+2)(\Delta-l-1)}g_{\Delta+3,l-1}\bigg]
	\end{dmath}
	Here we have presented the conformal partial wave in terms of conformal blocks and as one can observe, (\ref{eq:p}) is quite compact for the odd $l$. 
	\paragraph{}
	
	We can compare the above result of partial waves with that obtained for $\mathcal{N}=1$. The latter has been derived in  \cite{khandker2014n,berkooz2014bounds,fortin2011current}.
	The partial wave representing the four point correlation function in terms of the conformal blocks for odd spin is given by,
	\begin{equation}
		\mathcal{G}_{\Delta,l,odd}^{\mathcal{N}=1|JJ;\phi\phi^\dagger}=\lambda_{\mathcal{O}}\tilde{\lambda}_{\tilde{\mathcal{O}}} \bigg[-\frac{1}{2(\Delta+l+1)}g_{\Delta+1,l+1}+\frac{(l+2)(\Delta-l-2)}{2l(\Delta+l)(\Delta-l-1)}g_{\Delta+1,l-1}\bigg],
	\end{equation}
	while for even spin the expression becomes,
	\begin{equation}
		\mathcal{G}_{\Delta,l,even}^{\mathcal{N}=1|JJ;\phi\phi^\dagger}=\lambda_{\mathcal{O}}\tilde{\lambda}_{\tilde{\mathcal{O}}}\left[g_{\Delta,l}-\frac{1}{16}\frac{(\Delta-2)(\Delta+l)(\Delta-l-2)}{\Delta(\Delta+l+1)(\Delta-l-1)}g_{\Delta+2,l}\right].
	\end{equation}
	in (\ref{eq:p}), the ratio of the blocks $g_{\Delta+3,l+1}$ and $g_{\Delta+1,l+1}$ are nearly close to that of $\mathcal{N}=1$ even blocks. The ratio is not exactly matching with $\mathcal{N}=1$ because the number of supercharges is more in our case. 
	
	In order to examine the correctness of the expressions in (\ref{eq:p}), we perform the following consistency check. We consider the coefficients of the respective conformal blocks in the four point correlation functions of four $J$ \cite{li2020superconformal} and in the four point correlation functions of two chiral and two anti-chiral operators ($\langle\Phi\Phi^\dagger\Phi\Phi^\dagger\rangle$) \cite{fitzpatrick2014covariant} in ${\mathcal N}=2$ superconformal theories. 
Once we compare the square root of the product of those two coefficients with the coefficient of the conformal block given in (\ref{eq:p}) we get an exact match upto a sign and an overall factor. This implies that $\lambda_{\mathcal{O}}\tilde{\lambda}_{\tilde{\mathcal O}}$ appearing in (\ref{eq:p}) matches with the product of the 
analogous prefactors, which we denote by $\lambda_{\mathcal O}^{JJ}$ and $\lambda_{\mathcal O}^{\phi \phi^\dagger}$ respectively, appeared in the correlation functions of four $J$ and two chiral and two anti-chiral operators
 ($\langle\Phi\Phi^\dagger\Phi\Phi^\dagger\rangle$) upto an overall factor. In other words
	\begin{equation} \lambda_{\mathcal O} \tilde{\lambda}_{\tilde{\mathcal O}} = \frac{(2+l-\Delta)(\Delta - 1)(\Delta + l)}{8 \sqrt{2} (\Delta + 1)(\Delta - l)}\lambda_{\mathcal O}^{JJ} \lambda_{\mathcal O}^{\phi \phi^\dagger} \end{equation}
	
This shows that our result is consistent with the results obtained in \cite{fitzpatrick2014covariant,li2020superconformal}.
	
	
	\subsection{Result for $\mathcal{A}^\Delta_{0,0(\frac{l}{2},\frac{l}{2})}$ with $l$ even}
	In equation \ref{eq:8} we see that one of the three point functions contains two different kinds of superconformal invariants. Because of that, the OPE coefficients have two types of solutions, and the three point function contains more  tensor structures compared to its counterpart in the odd $l$. Inserting the superconformal projector, we obtain the superconformal partial wave for the even spin as,
	\begin{dmath}\label{eq:10}
		\mathcal{W}_{\mathcal{O}}\propto\int D[0,\bar{0}]\langle\mathcal{J}(1,\bar{1})\mathcal{J}(2,\bar{2})\mathcal{O}(0,\bar{0},\mathcal{S},\tilde{\mathcal{S}})\rangle\overleftrightarrow{\mathcal{D}}\langle\tilde{\mathcal{O}}(0,\bar{0},\mathcal{T},\tilde{\mathcal{T}})\Phi(3,\bar{3})\Phi^{\dagger}(4,\bar{4})\rangle \\
		=
		\frac{1}{(\langle 1\bar{2}\rangle\langle2\bar{1}\rangle)^{1-\frac{1}{4}(\Delta+l)}(\langle 3\bar{4}\rangle\langle 4\bar{3}\rangle)^{1-\frac{1}{4}(l-\Delta)}}\times
		\int D[0,\bar{0}]\frac{\mathcal{N'}^{full}_l}{(\langle 1\bar{0}\langle 2\bar{0}\rangle\langle0\bar{1}\rangle\langle0\bar{2}\rangle)^{\frac{1}{4}(l+\Delta)}(\langle 3\bar{0}\rangle\langle 4\bar{0}\rangle\langle 0\bar{3}\rangle\langle0\bar{4}\rangle)^{\frac{1}{4}(l-\Delta)}}
	\end{dmath}
	here,
	\begin{equation*}\mathcal{N'}^{full}_l =\mathcal{N'}^{(1)}_l+\mathcal{N'}^{(2)}_l+\mathcal{N'}^{(3)}_l+\mathcal{N'}^{(4)}_l+\mathcal{N'}^{(5)}_l+\mathcal{N'}^{(6)}_l
	\end{equation*}
	According to (\ref{eq:10}) 
	\begin{equation}\begin{split}
			\mathcal{N'}^{(1)}_l &= S_{-}^l\overleftrightarrow{\mathcal{D}}_l T_-^l (\lambda_0\tilde{\lambda}_0 + \lambda_2\tilde{\lambda}_0 z^2 + \lambda_3\tilde{\lambda}_0 \textbf w + \tilde{\lambda}_0\lambda_4 z^4 + \lambda_0\tilde{\lambda}_2 \tilde{z}^2 + \lambda_2\tilde{\lambda}_2 z^2 \tilde{z}^2 + \lambda_3 \tilde{\lambda}_2 \textbf w \tilde{z}^2 + \lambda_0 \tilde{\lambda}_4 \tilde{z}^4 )
			\\
			\mathcal{N'}^{(2)}_l &= l S_{-}^l \overleftrightarrow{\mathcal{D}}T_+ T_- ^{l-1} (\lambda_0 \tilde{\lambda}_1  \tilde{z} + \lambda_0 \tilde{\lambda}_3 \tilde{z}^3 + \lambda_2 \tilde{\lambda}_1 z^2 \tilde{z} + \lambda_3 \tilde{\lambda}_1 \textbf w \tilde{z}) 
			\\
			\mathcal{N'}^{(3)}_l &= S_+ S_-^{l-1} \overleftrightarrow {\mathcal{D}}_l T_-^l (\lambda_1 \tilde{\lambda}_0 z + \lambda_3 \tilde{\lambda}_0 {\textbf w} + \lambda_1 \tilde{\lambda}_2 z \tilde{z}^2)
			\\
			\mathcal{N'}^{(4)}_l &= l S_+ S_-^{l-1} \overleftrightarrow{D}_l T_+ T_-^{l-1} (\lambda_1 \tilde{\lambda}_1 z \tilde{z} + \lambda_1 \tilde{\lambda}_3 z \tilde{z}^3 )
			\\	
			\mathcal{N'}^{(5)}_l &= S_+^2 S_-^{l-2} \overleftrightarrow{D}_l T_-^l (\lambda_5 \tilde{\lambda}_0 + \lambda_5 \tilde{\lambda}_2 \tilde{z}^2 + \lambda_5 \tilde{\lambda}_4 \tilde{z}^4)
			\\	
			\mathcal{N'}^{(6)}_l &= l S_+^2 S_-^{l-2} \overleftrightarrow{D}_l T_+ T_-^{l-1} (\lambda_5 \tilde{\lambda}_1 \tilde{z} + \lambda_5 \tilde{\lambda}_3 \tilde{z}^3 )
	\end{split}\end{equation}
	The additional tensor structures are 
	\begin{equation}
		S_{+}^2S_{-}^{l-2}\overleftrightarrow{D}_lT_{-}^l=\frac{\mathcal{N}_l+\mathcal{M}_l}{8(l-1)\langle1\bar{2}\rangle^l\langle3\bar{4}\rangle^l}+(-1)^l(1\leftrightarrow2)+(-1)^l(3\leftrightarrow 4),
	\end{equation}
	\begin{equation}
		S_{+}^2S_{-}^{l-2}\overleftrightarrow{D}_lT_{+}T_{-}^{l-1}=\frac{\mathcal{N}_l+\mathcal{M}_l}{8l(l-1)\langle1\bar{2}\rangle^l\langle3\bar{4}\rangle^l}+(-1)^l(1\leftrightarrow2)-(-1)^l(3\leftrightarrow4),
	\end{equation}
	where,
	\begin{equation*}
		\mathcal{M}_l=\frac{1}{l!^2}(\partial_{\mathcal{S}}0\partial_{\mathcal{T}})^l(\mathcal{S}\bar{2}1\bar{0}3\bar{4}\mathcal{T})^{l-1}\mathcal{S}\bar{1}2\bar{0}3\bar{4}\mathcal{T}
	\end{equation*}
	All the above tensor structures will vanish when we set $\theta_{ext}=\bar{\theta}_{ext}=0$ but after the action of the partial derivatives $\partial_{\bar{0}}$ they may have non-zero contributions. As in the case of odd $l$,  after taking the derivatives the next step is to carry out the conformal integration. The necessary formulas for the integration are given in Appendix B.  Unlike the case of odd $l$, here the conformal integrations are mathematically more involved due to the presence of the derivative of the tensor structures. The final expression for the partial wave in terms of the conformal blocks for the even spin can be summarised as follows,
	\begin{equation}\label{leven}
		\mathcal{G}_{\Delta,l,even}^{\mathcal{N}=2|JJ;\phi\phi^{\dagger}}= c_0 g_{\Delta,l} + c_1 g_{\Delta+2,l+2} + c_2 g_{\Delta+2,l} + c_3 g_{\Delta+2,l-2} + c_4 g_{\Delta+4,l}.
	\end{equation}
	Unlike the odd $l$ case, it involves five conformal blocks and the coefficients are quite long, which we have given below.

		\begin{dmath*}\label{evenell}
		c_0=
		{\lambda}^{(1)}_{\mathcal{O}}\tilde{\lambda}_{\tilde{\mathcal{O}}}
			\end{dmath*}
	\begin{dmath*}
		c_1=\tilde{\lambda}_{\tilde{\mathcal{O}}}\left( \lambda_{\mathcal{O}}^{(1)}\frac{(\Delta+l)^2}{16(\Delta+l+1)(\Delta+l+3)}+\lambda_{\mathcal{O}}^{(2)}\frac{(\Delta+l)}{8(\Delta+l+3)}\right)
	\end{dmath*}
	\begin{dmath*}
		c_2=\tilde{\lambda}_{\tilde{\mathcal{O}}}\left(-{\lambda}_{\mathcal{O}}^{(1)}\frac{(2+l-\Delta)(\Delta+l)(\Delta+l\Delta+2)}{4l(\Delta+1)(l-\Delta)(2+l+\Delta)}-{\lambda}_{\mathcal{O}}^{(2)}\frac{(3+l)(2+l-\Delta)(2+\Delta)(\Delta+l)}{4l(l-\Delta)(\Delta + 1)(2+l+\Delta)}\right)
	\end{dmath*}
	\begin{dmath*}
		c_3=\tilde{\lambda}_{\tilde{\mathcal{O}}}\left(\lambda_{\mathcal O}^{(1)}\frac{(2+l-\Delta)(4+6l+5l^2+l^3-4\Delta-3l\Delta-l^2\Delta)}{16l(l-1)(l-\Delta-1)(l-\Delta+1)}+\lambda_{\mathcal O}^{(2)}\frac{(l+2)(l+3)(2+l-\Delta)}{8l(l-1)(l-\Delta-1)}\right)
	\end{dmath*}
	\begin{dmath*}
		c_4=\tilde{\lambda}_{\tilde{\mathcal{O}}}\left(\lambda_{\mathcal{O}}^{(1)}\frac{(2+l-\Delta)(\Delta+l)^2(4+2l-l\Delta+3\Delta^2+l\Delta^2-\Delta^3)}{256(l-\Delta-1)(l-\Delta+1)(\Delta+1)(\Delta+2)(\Delta+l+1)(\Delta+l+3)}+\lambda_{\mathcal{O}}^{(2)}\frac{(2+l)(3+l)(2+l-\Delta)\Delta(l+\Delta)}{64(l-\Delta-1)(l-\Delta+1)(\Delta+1)(\Delta+2)(\Delta+l+1)(\Delta+l+3)}\right)
	\end{dmath*}

	Here we have used a normalisation factor of $4(l-\Delta)^2(2+l+\Delta)^2$. As we have already mentioned, the contribution from $\langle \mathcal{J} \mathcal{J} {\mathcal {O}}\rangle$ involves two unknown overall coefficients, namely, ${\lambda}^{(1)}_{\mathcal{O}}$ and ${\lambda}^{(2)}_{\mathcal{O}}$ while the three point function $\langle\Phi\Phi^{\dagger}\mathcal{O}\rangle$ involves one unknown coefficient $\tilde{\lambda}_{\mathcal{O}}$. As has been discussed in \cite{li2020superconformal} the supershadow OPE coefficient $\tilde{\lambda}_{\tilde{\mathcal{O}}}$ is related to $\tilde{\lambda}_{\mathcal{O}}$.
		In the case of even $l$ too we can perform a similar consistency check by comparing the coefficients of the respective conformal blocks in the four point correlation functions of  four $J$ \cite{li2020superconformal} and in the four point correlation functions of two chiral and two anti-chiral operators ($\langle\Phi\Phi^\dagger\Phi\Phi^\dagger\rangle$) \cite{fitzpatrick2014covariant} as has been done in the case of odd $l$.
		 
	In the case of even $l$, the prefactors appearing in (\ref{evenell}) are $\tilde{\lambda}_{\tilde{\mathcal O}}\lambda^{(1)}_{\mathcal{O}}$ and $\tilde{\lambda}_{\tilde{\mathcal O}}\lambda^{(2)}_{\mathcal{O}}$. We denote the prefactor appearing in the correlation function of two chiral and two anti-chiral operators
	($\langle\Phi\Phi^\dagger\Phi\Phi^\dagger\rangle$) by $\lambda_{\mathcal O}^{\phi \phi^\dagger}$. For the correlation functions of four $J$ we use $\lambda_{\mathcal O}^{JJ(1)}$, $\lambda_{\mathcal O}^{JJ(2)}$ respectively, as in this case it involves two independent prefactors. 
	
	By comparing the coefficients of $g_{\Delta,l}$, $g_{\Delta+2,l+2}$, $g_{\Delta+2,l-2}$, $g_{\Delta+4,l}$, we find an exact match upto a potential sign and an overall factor. In this case, the prefactors are related through
	\begin{equation} 4(l-\Delta)^2(2+l+\Delta)^2 \lambda^{(m)}_{\mathcal O} \tilde{\lambda}_{\tilde{\mathcal O}} = \lambda_{\mathcal O}^{JJ(m)} \lambda_{\tilde{\mathcal O}}^{\phi \phi^\dagger},\quad m = 1 , 2. \end{equation}
	
	However, a similar comparison for $g_{\Delta+2,l}$ does not work because, the product of the coefficients of $g_{\Delta+2,l}$ from the correlation function of four $J$ and the correlation function of two chiral and two anti-chiral operators ($\langle\Phi\Phi^\dagger\Phi\Phi^\dagger\rangle$) does not form a perfect square. A possible reason could be that a number of multiplets contribute to the $g_{\Delta+2,l}$ term in the correlation function of four $J$ as explained in  \cite{li2020superconformal}.

	From the expressions in (\ref{leven}), one can see that $c_2, c_3$ and $c_4$ will vanish at unitarity bound i.e $\Delta=l+2$ and $c_0, c_1$ will be non zero. 
	 This vanishing may be a result of the fact that  at unitarity bound general long multiplet splits into 
	\begin{equation*}
		\mathcal{A}^{2+l}_{0,0(\frac{l}{2},\frac{l}{2})}=\hat{\mathcal{C}}_{0(\frac{l}{2},\frac{l}{2})}+\hat{\mathcal{C}}_{\frac{1}{2}(\frac{l-1}{2},\frac{l}{2})}+\hat{\mathcal{C}}_{\frac{1}{2}(\frac{l}{2},\frac{l-1}{2})}+\hat{\mathcal{C}}_{1(\frac{l-1}{2},\frac{l-1}{2})}
	\end{equation*}
	Out of these multiplets, only $\hat{\mathcal{C}}_{0(\frac{l}{2},\frac{l}{2})}$ will contribute to the four point correlator. Hence the conformal blocks $g_{\Delta,l}$ and $g_{\Delta+2,l+2}$ are the contribution from the multiplet $\hat{\mathcal{C}}_{0(\frac{l}{2},\frac{l}{2})}$. This is consistent with the decomposition of a $\hat{\mathcal{C}}_{0(\frac{l}{2},\frac{l}{2})}$ as given in \cite{dolan2003short}. A similar feature is exhibited by the coefficients in (\ref{eq:p}) for odd $l$ after an appropriate scaling of the overall coefficients $\lambda_{\mathcal{O}}$ and $\tilde{\lambda}_{\tilde{\mathcal{O}}}$.
	\paragraph{}
	We close this section with the following comments regarding analytic continuation of the result obtained above for general long multiplet $\mathcal{A}^{\Delta}_{0,0(\frac{l}{2}{\frac{l}{2}})}$. We have observed that in our case, we need to consider two exchange multiplets,	$\mathcal{A}^{2+l}_{0,0(\frac{l}{2},\frac{l}{2})}$ and $\hat{\mathcal{C}}_{0(\frac{l}{2},\frac{l}{2})}$ where the contribution due to the latter seems to be obtained by taking an appropriate limit.  However, when we analytically continue the superconformal block 	$\mathcal{G}_{\Delta,l,even}^{\mathcal{N}=2|JJ;\phi\phi^{\dagger}}$ to $\Delta=2,l=0$, as explained in \cite{li2020superconformal} the supershadow transformation for $\Delta=2,l=0$ may be pathological and unphysical term may arise from analytical continuation. It is not clear if there are unphysical terms in the analytical continuation $\Delta\rightarrow2+l$ of the conformal block for $l\geq0$. It may be interesting to investigate the issue further.
	
	\section{$\mathcal{N}=2$ superconformal block decompositions}
	The decomposition of $\mathcal{N}=2$ superconformal blocks into $\mathcal{N}=1$ helps us to check the consistency of our results. The decomposition occurs due to the decomposition of $\mathcal{N}=2$ multiplets into $\mathcal{N}=1$ multiplets. This relation has been used in \cite{khandker2014n,liendo2016stress,fortin2011current,poland2011bounds} for the consistency check of the $\mathcal{N}=1$ blocks. This agrees with the $\mathcal{N}=1$ decomposition of the $\mathcal{N}=2$ superconformal blocks considering the operators are global symmetry conserved current \cite{dolan2002superconformal}. This decomposition imposes non-trivial consistency checks and our results satisfy these as we will discuss below.
	\paragraph{}
	The decomposition of a general $\mathcal{N}=2$ superconformal long multiplet $\mathcal{A}^{\Delta}_{0,0(\frac{l}{2},\frac{l}{2})}$ has been solved in \cite{liendo2016stress}, which shows:
	\begin{equation}\label{6.1}
		\mathcal{A}^\Delta_{0,0(\frac{l}{2},\frac{l}{2})}\rightarrow A^{\Delta}_{r'=0(\frac{l}{2},\frac{l}{2})}+A^{\Delta+1}_{r'=0(\frac{l-1}{2},\frac{l-1}{2})}+A^{\Delta+1}_{r'=0(\frac{l+1}{2},\frac{l+1}{2})}+A^{\Delta+2}_{r'=0(\frac{l}{2},\frac{l}{2})}+A^{\Delta+1}_{r'=0(\frac{l-1}{2},\frac{l+1}{2})}+A^{\Delta+1}_{r'=0(\frac{l+1}{2},\frac{l-1}{2})},
	\end{equation}
	In this above equation, $\mathcal{A}$ is the $\mathcal{N}=2$ multiplet and $A$ are the $\mathcal{N}=1$ multiplets. The subindex $r'$=$\frac{2}{3}(2R+r)$ is the $U(1)_{r'}$ charge for $\mathcal{N}=1$ multiplets. We have not mentioned additional terms in (\ref{6.1}) with non-zero $r'$, as they do not contribute to the correlator $\langle JJ\phi\phi^\dagger\rangle$.  For odd $l$ the last two terms of the above equation (\ref{6.1}) do not contribute to the correlator $\langle JJ\phi\phi^\dagger\rangle$. Considering the first four terms, the decomposition implies,
	\begin{equation}\label{18}
		\mathcal{G}^{\mathcal{N}=2|JJ\phi\phi^\dagger}_{\Delta,l,odd}= a_0 \mathcal{G}^{\mathcal{N}=1}_{\Delta,l,odd} + a_1\mathcal{G}^{\mathcal{N}=1}_{\Delta+1,l-1,even}+a_2\mathcal{G}^{\mathcal{N}=1}_{\Delta+1,l+1,even} + a_3\mathcal{G}^{\mathcal{N}=1}_{\Delta+2,l,odd},
	\end{equation}
	Here $\mathcal{G}^{\mathcal{N}=1}_{\Delta,l,odd/even}$ are the $\mathcal{N}=1$ superconformal blocks. Using equation (\ref{18}), we expand the $\mathcal{N}=2$ superconformal blocks in terms of the bosonic conformal blocks as follows:
	\begin{equation} \label{18A}
		\mathcal{G}^{\mathcal{N}=2}_{\Delta,l,odd}\propto b_0 g_{\Delta+1,l+1} + b_1 g_{\Delta+1,l-1} + b_2 g_{\Delta+3,l+1} + b_3 g_{\Delta+3,l-1}.
	\end{equation}
	Comparing with (\ref{eq:p}) we can obtain the expressions of $\{b_0, b_1, b_2, b_3\}$ upto a constant.
	Expanding $\mathcal{G}^{\mathcal{N}=1}$ in terms of the conformal blocks $g$  and comparing that with (\ref{18A}) we can establish a linear relation (up to an overall multiplicative constant) between $\{a_0, a_1, a_2, a_3\}$ and $\{b_0, b_1, b_2, b_3\}$, as ${\mathbf a}. N = {\mathbf b}$. A straightforward computation shows that the matrix $N$ has a null eigenvector, and that gives rise to a relation among the coefficients $\{b_0, b_1, b_2, b_3\}$. Therefore, as in \cite{li2020superconformal} the coefficients $b_i$, which we have computed, need to satisfy a constraint, which is given as follows.
	\begin{equation}\label{eq:r}
		\frac{16 (\Delta + 1)(\Delta + l + 1)}{(\Delta - 1)(\Delta + l)}\left[ \frac{(l+2)(\Delta+l+3)}{l(\Delta + l +2)} b_2 + \frac{(\Delta - l+1)}{(\Delta - l)}
		b_3 \right] + \frac{(\Delta - l -2)(l + 2) (\Delta + l + 1) }{(\Delta - l - 1) l (\Delta + l) } b_0 + b_1 = 0 ,
	\end{equation}
	here $b_i$ are the coefficients of the conformal blocks presented in (\ref{eq:p}). Thus, the above equation (\ref{eq:r}) provides a nontrivial consistency check of our result for odd spin in equation (\ref{eq:p}) and our result satisfies the consistency check.
	\paragraph{}
	
	For even spin, the consistency condition arises through a similar analysis. As given by \cite{khandker2014n}, the $\mathcal{N}=2$ superconformal block can be expanded in terms of $\mathcal{N}=1$ blocks as follows,
	\begin{equation}\label{19}
		\mathcal{G}^{\mathcal{N}=2|JJ\phi\phi^\dagger}_{\Delta,l,even}= a^\prime_0 \mathcal{G}^{\mathcal{N}=1}_{\Delta,l,even} + a^\prime_1\mathcal{G}^{\mathcal{N}=1}_{\Delta+1,l-1,odd} + a^\prime_2 \mathcal{G}^{\mathcal{N}=1}_{\Delta+1,l+1,odd} + a^\prime_3 \mathcal{G}^{\mathcal{N}=1}_{\Delta+2,l,even},
	\end{equation}
	Substituting the expressions for the $\mathcal{N}=1$ blocks in terms of the conformal blocks we can write
	\begin{equation}
		\mathcal{G}^{\mathcal{N}=2}_{\Delta,l,even}\propto b^\prime_0 g_{\Delta , l } + b^\prime_1 g_{\Delta+2,l+2} + b^\prime_2 g_{\Delta+2,l} + b^\prime_3 g_{\Delta+2, l-2} +  b^\prime_4 g_{\Delta+4, l}.
	\end{equation}
	Clearly, $\{b^\prime_0, b^\prime_1, b^\prime_2, b^\prime_3, b^\prime_4\}$ can be identified with $\{c_0, c_1, c_2, c_3, c_4\}$ given in (\ref{leven}) upto an overall constant of proportionality. This time, since there are five parameters, $c_i$, which are linear combinations of four parameters $\{a^\prime_0, a^\prime_1, a^\prime_2, a^\prime_3 \}$, they satisfy the following constraint:
	\begin{equation}\begin{split}
			& \frac{\Delta- l -2}{\Delta - l -1} \left[ \frac{(\Delta - 2)(\Delta + l) }{16 \Delta (\Delta + l + 1)} c_0 + \frac{(l+3)(\Delta+l+3)}{(l+1)(\Delta+l+2)} c_1 \right]  + c_2 
			\\ &
			+ \frac{\Delta - l + 1}{\Delta - l } \left[ \frac{(l - 1 )(\Delta + l) }{(l+1) (\Delta + l + 1)} c_3 + \frac{16(\Delta + 2)(\Delta + l + 3) }{\Delta ( \Delta + l +2)} c_4 \right] = 0 .
	\end{split}\end{equation}
	The three point correlator $\langle \mathcal{J} \mathcal{J} {\mathcal O}\rangle$ involves $\lambda_{\mathcal O}^{(1)}$ and $\lambda_{\mathcal O}^{(2)}$, which are independent of each other. The expressions $c_i, i=0,1,2,3,4$ are linear in $\lambda_{\mathcal O}^{(1)}$ and $\lambda_{\mathcal O}^{(2)}$ and after putting the values of different $c_i$s the above consistency condition satisfies for each of the  $\lambda_{\mathcal O}^{(i)}$s.

	\section{Discussion}
	
	In this work, we have evaluated the superconformal partial wave $\langle J J \phi\phi^\dagger\rangle$ using the superembedding space. The three point correlators $\langle \mathcal{J}\mathcal{J} {\mathcal O}\rangle$ and $\langle \Phi\Phi^\dagger{\mathcal O}\rangle$ can be expanded in terms of the superconformal invariants and the tensor structures. The coefficients are determined using the relevant equations and ${\mathbb Z}_2$ reflection symmetry. Then using shadow formalism we obtain the four point correlator as a sum over conformal blocks. This mixed correlator has non-zero contribution only for the general long multiplet and for $\hat{\mathcal C}_{0,(\frac{l}{2},\frac{l}{2})}$. We have discussed the contribution for the general long multiplet. It is to be checked whether the contribution for the latter can be obtained as a limit of the former. Compared with the ${\mathcal N}=1$ superconformal blocks, one can find a consistency check for odd $l$. For even $l$ there are two parameters,  $\lambda_{\mathcal O}^{(1)}$ and $\lambda_{\mathcal O}^{(2)}$, and it turns out for each parameter there is a consistency check. We have found that our result satisfies all the consistency checks. Considering the fact that the coefficients of the conformal blocks in the expressions of the four point correlators are involved it is quite satisfactory.
	
	The present analysis deals with the mixed correlator in the  ${\mathcal N}=2$ theory. The power of mixed correlators in the context of three dimensional bootstrap has been elaborated in \cite{kos2014bootstrapping}. A numerical study of the bootstrap of a mixed four-point correlator for chiral and real scalars in $D=4$  ${\mathcal N}=1$ has been done in \cite{li2017bootstrapping}. The present model provides part of the ingredients necessary for the bootstrap analysis of a mixed correlator in $D=4$,  ${\mathcal N}=2$ theory. A similar analysis of the correlation function in the crossed channel, $\langle J  \phi J\phi^\dagger\rangle$ is also required for the bootstrap analysis.
	
	It has been found that the critical three dimensional Ising model lives at the kink of the bound on the CFT data, \cite{el2012solving,el2014solving,kos2014bootstrapping}. Plausibly, a CFT containing a minimum spectrum of operators to satisfy the crossing relation lives at kink. When one approaches the kink along the boundary of the allowed region, squared OPE coefficients of certain operators vanish, which can be interpreted as two different solution branches meeting at the kink \cite{poland2019conformal,el2014solving,Li:2017kck} and leads to decoupling of certain operators. One may explore the possibility of a similar scenario for the mixed correlator of $4D$ ${\mathcal N}=2$.  Since stress tensor operator is universal in any local CFT, it is natural to consider correlators involving stress tensor multiplet. However, it is not clear which theories are relevant to this possibility. But as mentioned in \cite{beem2016n}, $4D$ ${\mathcal N}=2$ superconformal QCD with gauge group $SU(2)$, $N_f=4$ may deserve further study \cite{Chester:2022sqb}.
	
 Though the four point correlator of stress-tensor multiplet for ${\mathcal N}=2$ has been discussed in \cite{li2020superconformal}, a similar analysis for ${\mathcal N}=1$ theory has not been done. A further generalization of the present work is to analyze the mixed correlators involving the vector and higher spin currents in superconformal field theories. In this context, it may be mentioned that higher-spin supercurrents in $D=4$ ${\mathcal N}=1$ superconformal theory has been studied in \cite{buchbinder2022three}. We hope to report on some of these in the future.
	
	\section*{Acknowledgement}
	A preliminary version of this work was presented at the XXV DAE-BRNS High Energy Physics Symposium 2022, IISER Mohali. SR is thankful to the local organizing committee for the hospitality during the stay at IISER Mohali.

	\appendix
	\section{Some invariants and tensor structures}
	\subsection{Superconformal invariants and its derivatives}
	Two independent invariants in $\mathcal{N}=2$ superembedding space are $z$ and $\textbf w$ which are constructed by taking the 2-point traces of three points $(1,2,0)$:
	\begin{equation}
		z=\frac{\langle\bar{1}2\rangle\langle\bar{2}0\rangle\langle\bar{0}1\rangle-\langle\bar{2}1\rangle\langle\bar{1}0\rangle\langle\bar{0}2\rangle}{\langle\bar{1}2\rangle\langle\bar{2}0\rangle\langle\bar{0}1\rangle+\langle\bar{2}1\rangle\langle\bar{1}0\rangle\langle\bar{0}2\rangle}
	\end{equation}
	and the six point trace will produce 
	\begin{equation}
		\textbf w=\frac{4\langle\bar{1}2\bar{0}1\bar{2}0\rangle}{(\langle\bar{1}2\rangle\langle\bar{2}1\rangle\langle\bar{1}0\rangle\langle\bar{0}1\rangle\langle\bar{0}2\rangle\langle\bar{2}0\rangle)^{\frac{1}{2}}}+1.
	\end{equation}
	By exchanging the coordinates by $1\leftrightarrow3$ and $2\leftrightarrow4$ we obtain other superconformal invariants in $\mathcal{N}=2$ superembedding space $\bar{z}$ and $\bar{\textbf w}$. All the superconformal invariants are nilpotent and vanish after setting external Grassmann variables to zero, which reads
	\begin{eqnarray*}
		z^5= \textbf w^3=0\\
		z|_{\theta_{ext}=0}=\textbf w|_{\theta_{ext}}=0.
	\end{eqnarray*}	
	$z$ or $w$ will vanish unless the derivatives $\partial_{\bar{0}}$ act on them. The derivatives on $z$ are as follows: 
	\begin{equation}\label{A1}
		\partial_{\bar{0}}^mz=(z^2-1)\left(\frac{X_1^m}{X_{1\bar{0}}}-\frac{X_2^m}{X_{2\bar{0}}}\right), \hspace{1cm} \partial_{\bar{0}}^2z=2z(z^2-1)\frac{X_{12}}{X_{1\bar{0}}X_{2\bar{0}}}
	\end{equation}
	\begin{equation}\label{A2}
		\partial_{\bar{0}}^mz^2=2z(z^2-1)\left(\frac{X_1^m}{X_{1\bar{0}}}-\frac{X_2^m}{X_{2\bar{0}}}\right), \hspace{1cm} \partial_{\bar{0}}^2z^2=(6z^2-2)(z^2-1)\frac{X_{12}}{X_{1\bar{0}}X_{2\bar{0}}}
	\end{equation}
	\begin{equation} \label{A3}
		(\partial_{\bar{0}}z).(\partial_{\bar{0}}\tilde {z})|_{\bar{0}=0}=-\frac{1}{2}\left(\frac{X_{13}}{X_{1\bar{0}}X_{3\bar{0}}}-(1\leftrightarrow2)-(3\leftrightarrow4)\right)
	\end{equation}
	\begin{equation}\label{A4}
		\partial_{\bar{0}}^m\partial_{\bar{0}}^nz=2(z^2-1)\left(z\left(\frac{X_1^m}{X_{1\bar{0}}}-\frac{X_2^m}{X_{2\bar{0}}}\right)\left(\frac{X_1^n}{X_{1\bar{0}}}-\frac{X_2^n}{X_{2\bar{0}}}\right)+\left(\frac{X_1^mX_1^n}{X_{1\bar{0}}X_{1\bar{0}}}-\frac{X_2^mX_2^n}{X_{2\bar{0}}X_{2\bar{0}}}\right)\right)
	\end{equation}
	These are also anti-symmetric under the coordinate interchange $1\leftrightarrow2$. The higher order derivatives on $z$ can be obtained using the same procedure and also for the $z^2, z^3 \text{and}  z^4$. If the degree of $z$ is larger than the order of the derivative, the ultimate contribution after setting $\theta=\bar{\theta}=0$ will become zero. The same will be applicable for $\textbf w$. The derivatives on $\textbf w$ are as follows:
	\begin{equation} \label{A5}
		\partial_{\bar{0}}^m\textbf w=\frac{16\langle\bar{1}2\bar{\Gamma}^m1\bar{2}0\rangle}{(X_{1\bar{0}}X_{0\bar{1}}X_{2\bar{0}}X_{0\bar{2}}X_{2\bar{1}}X_{1\bar{2}})^{\frac{1}{2}}}+(\textbf w-1)\left(\frac{X_1^m}{X_{1\bar{0}}}+\frac{X_2^m}{X_{2\bar{0}}}\right),
	\end{equation}
	Unlike $z$, $\textbf w$ is symmetric under the permutation of $1\leftrightarrow2$ and the higher order derivatives are :
	\begin{equation} \label{A6}
		\partial_{\bar{0}}^2\textbf w=-(\textbf w-1)\frac{X_{12}}{X_{1\bar{0}}X_{2\bar{0}}},
	\end{equation}
	\begin{equation} \label{A7}
		\partial_{\bar{0}}^m\partial_{\bar{0}}^n\textbf w=\partial_{\bar{0}}^m\textbf w\left(\frac{X_1^n}{X_{1\bar{0}}}+\frac{X_2^n}{X_{2\bar{0}}}\right)+\partial_{\bar{0}}^n\textbf w\left(\frac{X_1^m}{X_{1\bar{0}}}+\frac{X_2^m}{X_{2\bar{0}}}\right)+(\textbf w-1)\left(\frac{X_1^m}{X_{1\bar{0}}}-\frac{X_2^m}{X_{2\bar{0}}}\right)\left(\frac{X_1^n}{X_{1\bar{0}}}-\frac{X_2^n}{X_{2\bar{0}}}\right),
	\end{equation}
	\begin{equation} \label{A8}
		\partial_{\bar{0}}^m\partial_{\bar{0}}^2\textbf w=-\partial_{\bar{0}}^m\textbf w\frac{X_{12}}{X_{1\bar{0}}X_{2\bar{0}}}-2 (\textbf w-1)\frac{X_{12}}{X_{1\bar{0}}X_{2\bar{0}}}\left(\frac{X_1^m}{X_{1\bar{0}}}+\frac{X_2^m}{X_{2\bar{0}}}\right)
	\end{equation}
	\begin{equation} \label{A9}
		\partial_{\bar{0}}^2\partial_{\bar{0}}^2\textbf w=9(\textbf w-1)\left(\frac{X_{12}}{X_{1\bar{0}}X_{2\bar{0}}}\right)^2
	\end{equation}
	The bare $\textbf w$ always gives $0$ after setting $\theta_{ext}=0$. To get the non-zero values, the order of $\textbf w$ must be less than or equal to the degree of the partial derivatives. Setting $\theta_{ext}=0$ we obtain,
	\begin{equation} \label{A10}
		\partial_{\bar{0}}^m\textbf w|_{\theta_{ext}=0}=-\frac{X_{12}}{X_{1\bar{0}}X_{2\bar{0}}}X_{0}^m.
	\end{equation}
	On the other hand, $D_l$ is invariant under the exchanges of $1\leftrightarrow2$ and $3\leftrightarrow4$ and the derivatives reads
	\begin{equation} \label{A11}
		\partial_{\bar{0}}^m\frac{1}{D_l}=\frac{1}{D_l}\left[\left(\frac{\Delta+l}{2}\right)\left(\frac{X_1^m}{X_{1\bar{0}}}+\frac{X_2^m}{X_{2\bar{0}}}\right) - \left(\frac{\Delta-l}{2}\right)\left(\frac{X_3^m}{X_{3\bar{0}}}+\frac{X_4^m}{X_{4\bar{0}}}\right)\right],
	\end{equation}
	\begin{equation} \label{A12}
		\partial_{\bar{0}}^2\frac{1}{D_l}=-\frac{1}{D_l}\left[\left(\frac{\Delta+l}{2}\right)^2\frac{X_{12}}{X_{1\bar{0}}X_{2\bar{0}}}+\left(\frac{\Delta-l}{2}\right)^2\frac{X_{34}}{X_{3\bar{0}}X_{4\bar{0}}}-\left(\frac{\Delta+l}{2}\right)\left(\frac{\Delta-l}{2}\right)\left(
		\Omega_{+}\right) 
		\right]
	\end{equation}
	where 
	\begin{equation*}
		\Omega_{+}=\left(\frac{X_{13}}{X_{1\bar{0}}X_{3\bar{0}}}+\text{even permutations}\right).
	\end{equation*}
	\subsection{Tensor structures and its derivatives}
	In this section, we will deal with the single order partial derivatives and second order partial derivatives on tensor structures, because of different permutations of $\mathcal{N}_l$ and $\mathcal{L}_l$ the first order derivatives on the tensor structures will give rise to new tensor structures \cite{li2016most} which consists of P0, P1, P2, P3, R0, R1, R2, R3.\\ In the equation, we can write them as, 
	\begin{eqnarray}
		(\partial_S0\partial_T)^l(S21034T)^{l-1}(X_{10}S234T+X_{20}S134T+X_{10}S243T+X_{20}S143T)\\
		(\partial_S0\partial_T)^l(S21034T)^{l-1}(X_{10}S234T+X_{20}S134T-X_{10}S243T-X_{20}S143T)\\
		(\partial_S0\partial_T)^l(S21034T)^{l-1}(X_{10}S234T-X_{20}S134T+X_{10}S243T-X_{20}S143T)\\
		(\partial_S0\partial_T)^l(S21034T)^{l-1}(X_{10}S234T-X_{20}S134T-X_{10}S243T+X_{20}S143T).
	\end{eqnarray}
	As per the convention 
	\begin{eqnarray*}
		P0=(X_{10}S234T+X_{20}S134T+X_{10}S243T+X_{20}S143T)\\
		P1=(X_{10}S234T+X_{20}S134T-X_{10}S243T-X_{20}S143T)\\
		P2=(X_{10}S234T-X_{20}S134T+X_{10}S243T-X_{20}S143T)\\
		P3=(X_{10}S234T-X_{20}S134T-X_{10}S243T+X_{20}S143T)
	\end{eqnarray*}
	Similarly, we reduce the tensors structures to compact forms proportional to $N_l$,
	\begin{eqnarray}
		(\partial_S0\partial_T)^l(S21034T)^{l-1}(X_{30}S214T+X_{40}S213T+X_{30}S124T+X_{40}S123T)\\
		(\partial_S0\partial_T)^l(S21034T)^{l-1}(X_{30}S214T+X_{40}S213T-X_{30}S124T-X_{40}S123T)\\
		(\partial_S0\partial_T)^l(S21034T)^{l-1}(X_{30}S214T-X_{40}S213T+X_{30}S124T-X_{40}S123T)\\
		(\partial_S0\partial_T)^l(S21034T)^{l-1}(X_{30}S214T-X_{40}S213T-X_{30}S124T+X_{40}S123T).
	\end{eqnarray}
	Again as per  convention,
	\begin{eqnarray*}
		R0=(X_{30}S214T+X_{40}S213T+X_{30}S124T+X_{40}S123T)\\
		R1=(X_{30}S214T+X_{40}S213T-X_{30}S124T-X_{40}S123T)\\
		R2=(X_{30}S214T-X_{40}S213T+X_{30}S124T-X_{40}S123T)\\
		R3=(X_{30}S214T-X_{40}S213T-X_{30}S124T+X_{40}S123T)
	\end{eqnarray*}
	The tensor structure related to $SPT$ and $SRT$ are :
	\begin{equation}
		P_l=\frac{1}{l!^2}(\partial_{\mathcal{S}}0\partial_{\mathcal{T}})^l(S\bar{2}1\bar{0}3\bar{4}T)^{l-1}\times(X_{10}S\bar{2}3\bar{4}T+X_{20}S\bar{1}3\bar{4}T-X_{10}S\bar{2}4\bar{3}T-X_{20}S\bar{1}4\bar{3}T),
	\end{equation}
	and,
	\begin{equation}
		R_l=\frac{1}{l!^2}(\partial_{\mathcal{S}}0\partial_{\mathcal{T}})^l(S\bar{2}1\bar{0}3\bar{4}T)^{l-1}\times(X_{30}S\bar{2}1\bar{4}T+X_{40}S\bar{2}1\bar{3}T-X_{30}S\bar{1}2\bar{4}T-X_{40}S\bar{1}2\bar{3}T).
	\end{equation}
	\section{List of Conformal Integrations}
	\subsection{Integration related to $\mathcal{N}_l$ } In this section, we will discuss how to perform the conformal integration. We will provide the formulas used in this work to evaluate the conformal integrations. The conformal integration related to $N_l$, or Gegenbauer polynomial $C^{(1)}_l(t_0)$ are given by,
	\begin{equation*}
		\int_M D^4X_0\frac{(-1)^lC^{(1)}_l(t_0)}{X_{10}^{\frac{\Delta+r}{2}}X_{20}^{\frac{\Delta-r}{2}}X_{30}^{\frac{\tilde{\Delta}+\tilde{r}}{2}}X_{40}^{\frac{\tilde{\Delta}-\tilde{r}}{2}}}=\xi_{\Delta,\tilde{\Delta},\tilde{r},l}\bigg(\frac{X_{14}}{X_{34}}\bigg)^{\frac{\tilde{r}}{2}}\bigg(\frac{X_{24}}{X_{14}}\bigg)^{\frac{r}{2}}X_{12}^{-\frac{\Delta}{2}}X_{34}^{-\frac{\tilde{\Delta}}{2}}g_{\Delta,l}^{r,\tilde{r}}(u,v).
	\end{equation*}
	In which $r=\Delta_1-\Delta_2$,$\tilde{r}=\Delta_3-\Delta_4$ and 
	\begin{equation*}
		\xi_{\Delta,\tilde{\Delta},\tilde{r},l}=\frac{\pi^2\Gamma(\Delta+l-1)\Gamma(\frac{\Delta-\tilde{r}+l}{2})\Gamma(\frac{\Delta+\tilde{r}+l}{2})}{(2-\Delta)\Gamma(\Delta+l)\Gamma(\frac{\tilde{\Delta}-\tilde{r}+}{2})\Gamma(\frac{\tilde{\Delta}+\tilde{r}+l}{2})}
	\end{equation*}
	The closed form structure of the conformal block $g^{r,\tilde{r}}_{\Delta,l}(u,v)$ is,
	\begin{equation*}
		g^{r,\tilde{r}}_{\Delta,l}(u,v)=\frac{\rho\bar{\rho}}{\rho-\bar\rho}[k_{\Delta+l}(\rho)k_{\Delta-l-2}(\bar{\rho})-(\rho \leftrightarrow{\bar{\rho}})]
	\end{equation*} and
	\begin{equation*}
		k_\beta(x)={x^{\frac{\beta}{2}}}_2F_1\bigg(\frac{\beta-r}{2},\frac{\beta+\tilde{r}}{2},\beta,x\bigg),
	\end{equation*} 
	where the conformal invariants $u,v$ are $\rho\bar{\rho}$ and $(1-\rho)(1-\bar{\rho})$ respectively.
	\begin{equation}
		\int D^4X_0\frac{X_{12}}{X_{10}X_{20}}\frac{N_l}{D_l}\bigg|_{\bar{0}=0}=\frac{2^{-6l}\xi_{\Delta+2,2-\Delta,0,l}}{X^{\frac{\Delta-l}{2}}_{12}X^{-\frac{\Delta+l}{2}}_{34}}g_{\Delta+2,l}
	\end{equation}
	\begin{equation}
		\int D^4X_0\frac{X_{34}}{X_{30}X_{40}}\frac{N_l}{D_l}\bigg|_{\bar{0}=0}=\frac{2^{-6l}\xi_{\Delta,4-\Delta,0,l}}{X^{\frac{\Delta-l}{2}}_{12}X^{-\frac{\Delta+l}{2}}_{34}}g_{\Delta,l}
	\end{equation}
	\begin{equation}
		\int D^4X_0\frac{N_{l-1}}{D_l}\bigg|_{\bar{0}=0}=\frac{2^{-6(l-1)}\xi_{\Delta+1,3-\Delta,0,l-1}}{X^{\frac{\Delta-l}{2}}_{12}X^{-\frac{\Delta+l}{2}}_{34}}g_{\Delta+1,l-1}
	\end{equation}
	
	\paragraph{}
	\textbf{Some conformal integrations related to $\Omega_x$:}
	Before listing out the integration with $\Omega_x$ $(x\in +,-,A,B)$, we mention 
	\begin{equation}
		\Omega_{+}=\frac{X_{13}}{X_{10}X_{30}}+\frac{X_{23}}{X_{20}X_{30}}+\frac{X_{14}}{X_{10}X_{40}}+\frac{X_{24}}{X_{20}X_{40}}
	\end{equation}
	\begin{equation}
		\Omega_{-}=\frac{X_{13}}{X_{10}X_{30}}-\frac{X_{23}}{X_{20}X_{30}}-\frac{X_{14}}{X_{10}X_{40}}+\frac{X_{24}}{X_{20}X_{40}}
	\end{equation}
	\begin{equation}
		\Omega_{A}=\frac{X_{13}}{X_{10}X_{30}}-\frac{X_{23}}{X_{20}X_{30}}+\frac{X_{14}}{X_{10}X_{40}}-\frac{X_{24}}{X_{20}X_{40}}
	\end{equation}
	\begin{equation}
		\Omega_{A}=\frac{X_{13}}{X_{10}X_{30}}+\frac{X_{23}}{X_{20}X_{30}}-\frac{X_{14}}{X_{10}X_{40}}-\frac{X_{24}}{X_{20}X_{40}}
	\end{equation}
	$\Omega_x$ are anti-symmetric under coordinate exchange of $1\leftrightarrow2$ and $3\leftrightarrow4$. The integrations are listed below.
	\begin{equation}
		\int D^4X_0 \frac{X_{12}}{X_{10}X_{20}}\Omega_{A}\frac{N_l}{D_l}=\int D^4X_0 \frac{X_{12}}{X_{10}X_{20}}\Omega_{B}\frac{N_l}{D_l}=0
	\end{equation}
	\begin{equation}
		\int D^4X_0 \frac{X_{34}}{X_{30}X_{40}}\Omega_{A}\frac{N_l}{D_l}=\int D^4X_0 \frac{X_{34}}{X_{30}X_{40}}\Omega_{B}\frac{N_l}{D_l}=0
	\end{equation}
	\begin{equation}
		\int D^4X_0\Omega_{+}\frac{N_l}{D_l}=\frac{2^{-6l}\xi_{\Delta+1,3-\Delta,1,l}}{X^{\frac{\Delta-l}{2}}_{12}X^{-\frac{\Delta+l}{2}}_{34}}\bigg[4g_{\Delta,l}+\frac{(\Delta+l)(\Delta-l-2)}{4(\Delta+l+1)(\Delta-l-1)}g_{\Delta+2,l}\bigg]
	\end{equation}
	\begin{equation}
		\int D^4X_0\Omega_{-}\frac{N_l}{D_l}=\frac{2^{-6l}\xi_{\Delta+1,3-\Delta,1,l}}{X^{\frac{\Delta-l}{2}}_{12}X^{-\frac{\Delta+l}{2}}_{34}}\bigg[\frac{\Delta+l}{\Delta+l+1}g_{\Delta+1,l+1}+\frac{(\Delta-l-2)}{(\Delta-l-1)}g_{\Delta+1,l-1}\bigg]
	\end{equation}
	\begin{dmath}
		\int D^4X_0 \Omega^{2}_{-}\frac{N_l}{D_l}=\frac{2^{-6l} \xi_{\Delta+2,2-\Delta,2,l}}{X^{\frac{\Delta-l}{2}}_{12}X^{-\frac{\Delta+l}{2}}_{34}}\left(\frac{16((\Delta-l)(\Delta+l)-2l)}{(\Delta+l+2)(\Delta-l)}g_{\Delta,l}+\frac{2(\Delta-l-2)(\Delta+l)}{(\Delta+l+2)(\Delta-l)}g_{\Delta+2,l}\\+\frac{((\Delta-l)(\Delta+l)-2l)(\Delta-l-2)(\Delta+l)}{16(\Delta+l+3)(\Delta+l+1)(\Delta-l+1)(\Delta-l-1)}g_{\Delta+4,l}-\frac{\Delta+l}{(\Delta+l+3)(\Delta-l)(\Delta+l+1)}g_{\Delta+2,l+2}\\-\frac{(\Delta-l-2)}{(\Delta-l-1)(\Delta-l+1)(\Delta+l+2)}g_{\Delta+2,l-2}\right)
	\end{dmath}
	\begin{dmath}
		\int D^4X_0 \Omega_{-}^2\frac{N_l}{D_l}=\frac{2^{-6l}\xi_{\Delta+2,2-\Delta,2,l}}{X^{\frac{\Delta-l}{2}}_{12}X^{-\frac{\Delta+l}{2}}_{34}}\left(\frac{2(\Delta-l-2)(\Delta+l)}{(\Delta+l+2)(\Delta-l)}g_{\Delta+2,l}\\+\frac{(\Delta+l)(\Delta-l-1)}{(\Delta+l+3)(\Delta-l)}g_{\Delta+2,l+2}+\frac{(\Delta-l-2)(\Delta+l+1)}{(\Delta-l+1)(\Delta+l+2)}g_{\Delta+2,l-2}\right)
	\end{dmath}
	\begin{dmath}
		\int D^4X_0 \Omega_{A}^2\frac{N_l}{D_l}=\frac{2^{-6l}\xi_{\Delta+2,2-\Delta,2,l}}{X^{\frac{\Delta-l}{2}}_{12}X^{-\frac{\Delta+l}{2}}_{34}}\left(-\frac{\Delta(\Delta-l-2)(\Delta+l)}{8(\Delta+l+3)(\Delta+l+1)(\Delta-l+1)(\Delta-l-1)}g_{\Delta+4,l}+\frac{\Delta+l}{(\Delta+l+3)(\Delta-l)}g_{\Delta+2,l+2}+\frac{\Delta-l-2}{(\Delta-l+1)(\Delta+l+2)}g_{\Delta+2,l-2}\right)
	\end{dmath}
	\begin{dmath}
		\int D^4X_0\Omega_{B}^2\frac{N_l}{D_l}=\frac{2^{-6l}\xi_{\Delta+2,2-\Delta,2,l}}{X^{\frac{\Delta-l}{2}}_{12}X^{-\frac{\Delta+l}{2}}_{34}}\left(\frac{32\Delta}{(\Delta+l+2)(\Delta-l)}g_{\Delta,l}-\frac{(\Delta-l-1)(\Delta+l)}{(\Delta+l+3)(\Delta-l)(\Delta+l+1)}g_{\Delta+2,l+2}-\frac{(\Delta-l-2)(\Delta+l+1)}{(\Delta-l-1)(\Delta-l+1)(\Delta+l+2)}g_{\Delta+2,l-2}\right)
	\end{dmath}
	\begin{dmath}
		\int D^4 X_0(\Omega_{A}\Omega_{B}+\Omega_{+}\Omega_{-})\frac{N_l}{D_l}=\frac{2^{-6l}\xi_{\Delta+2,2-\Delta,2,l}}{X^{\frac{\Delta-l}{2}}_{12}X^{-\frac{\Delta+l}{2}}_{34}}\left(\frac{4(\Delta+l)}{\Delta+l+2}g_{\Delta+1,l+1}+4\frac{\Delta-l-2}{\Delta-l}g_{\Delta+1,l-1}+\frac{(\Delta-l-2)(\Delta+l)^2}{4(\Delta-l)(\Delta+l+1)(\Delta+l+3)}g_{\Delta+3,l+1}+\frac{(\Delta-l-2)^2(\Delta+l)}{4(\Delta-l+1)(\Delta-l-1)(\Delta+l+2)}g_{\Delta+3,l-1}\right)
	\end{dmath}
	\subsection{Conformal integration with (P0, P1, P2, P3) and (R0, R1, R2, R3)}
	
	For $\mathcal {N}=2$ theories, the action of double derivatives $\partial_{0}^n\partial_{0}^m$ on the tensor structures will lead to higher order tensor structures. For higher order tensor structures, conformal integrations are not very straightforward. One way to simplify the higher order tensor structures is to reduce them to first order tensor structures, wherever possible. In fact, one can show that $P3=R3=8S21034T$, while $P2$ and $R2=0$ vanish (see \cite{li2016most} for $P0$, $R0$ and other details). The tensor structures $P1R1$, $P1^2$ and $R1^2$ cannot be reduced further. We have closely followed \cite{li2020superconformal} for computing our results and most of the integrations are available in the appendix of the above mentioned paper. 
	\begin{dmath}
		\int D^4X_0 \frac{X_{12}}{X_{10}X_{20}}\Omega_{A}\frac{P_l}{D_l}=\frac{2^{3-6l}\xi_{\Delta+3,1-\Delta,0,l-1}}{X_{12}^{\frac{1}{2}(\Delta-l)}X_{34}^{-\frac{1}{2}(\Delta+l)}}\left(\frac{4(\Delta-l+1)}{\Delta l(l-\Delta)}g_{\Delta+2,l}-\frac{4(\Delta+1)(l+1)}{\Delta l(\Delta+l)}g_{\Delta+2,l-2}\\+\frac{\Delta+l+2}{4(\Delta+l+1)(\Delta+l+3)}g_{\Delta+4,l}\right)
	\end{dmath}
	\begin{dmath}
		\int D^4X_0 \frac{X_{12}}{X_{10}X_{20}}\Omega_{B}\frac{P_l}{D_l}=\frac{2^{3-6l}\xi_{\Delta+3,1-\Delta,0,l-1}}{X_{12}^{\frac{1}{2}(\Delta-l)}X_{34}^{-\frac{1}{2}(\Delta+l)}}\left(-\frac{1}{\Delta l}g_{\Delta+3,l-1}-\frac{16(\Delta+1)(l+1)(\Delta-l+1)}{\Delta l(\Delta+l)(\Delta-l)}g_{\Delta+1,l-1}\\+\frac{(\Delta+l+2)(\Delta-l+1)}{(\Delta-l)(\Delta+l+1)(\Delta+l+3)}g_{\Delta+3,l+1}\right)
	\end{dmath}
	\begin{dmath}
		\int D^4X_0 \frac{X_{34}}{X_{30}X_{40}}\Omega_{A}\frac{R_l}{D_l}=\frac{2^{3-6l}\xi_{\Delta+1,3-\Delta,0,l-1}}{X_{12}^{\frac{1}{2}(\Delta-l)}X_{34}^{-\frac{1}{2}(\Delta+l)}}\left(-\frac{\Delta+l}{(\Delta+l+1)(\Delta-l-2)}g_{\Delta+1,l+1}\\+\frac{1}{\Delta l}g_{\Delta+1,l-1}\\+\frac{(\Delta-1)(l+1)(\Delta+l)(\Delta-l)}{16\Delta l(\Delta+l+1)(\Delta-l-1)(\Delta-l+1)}g_{\Delta+3,l-1}\right)
	\end{dmath}
	\begin{dmath}
		\int D^4X_0 \frac{X_{34}}{X_{30}X_{40}}\Omega_{B}\frac{R_l}{D_l}=\frac{2^{3-6l}\xi_{\Delta+1,3-\Delta,0,l-1}}{X_{12}^{\frac{1}{2}(\Delta-l)}X_{34}^{-\frac{1}{2}(\Delta+l)}}
		\left(-\frac{4}{\Delta-l-2}g_{\Delta,l}+\frac{\Delta+l}{4\Delta l(\Delta+l+1)}g_{\Delta+2,l}\\+\frac{(\Delta-1)(l+1)(\Delta-l)}{4\Delta l(\Delta-l-1)(\Delta-l+1)}g_{\Delta+2,l-2}\right)
	\end{dmath}
	The conformal integration  involving $P1^2$ and $R1^2$ are\cite{li2020superconformal}
	\begin{dmath}
		\int D^4X_0\left(\frac{X_{12}}{X_{10}X_{20}}\right)^2\frac{1}{D_l}\frac{l(l-1)}{(l!)^2}(\partial_{S}0\partial_{T})^l(S\bar{2}1\bar{0}3\bar{4}T)^{l-2}P1^2=\frac{2^{6(l-1)}\xi_{\Delta+2,2-\Delta,0,l-2}}{X_{12}^{\frac{1}{2}(\Delta-l)}X_{34}^{-\frac{1}{2}(\Delta+l)}}\left(\frac{l(l+1)}{\Delta+l+1}g_{\Delta+2,l+2}\\-\frac{l(l-1)\Delta(\Delta+l)(\Delta+l+2)}{16(\Delta+2)(\Delta+l+1)^2(\Delta+l+3)}g_{\Delta+4,l}\right)
	\end{dmath}
	\begin{dmath}
		\int D^4X_0\left(\frac{X_{34}}{X_{30}X_{40}}\right)^2\frac{1}{D_l}\frac{l(l-1)}{(l!)^2}(\partial_{S}0\partial_{T})^l(S\bar{2}1\bar{0}3\bar{4}T)^{l-2}R1^2=\frac{2^{6(l-1)}\xi_{\Delta+2,2-\Delta,0,l-2}}{X_{12}^{\frac{1}{2}(\Delta-l)}X_{34}^{-\frac{1}{2}(\Delta+l)}}\left(l(l-1)\frac{\Delta+l+2}{(\Delta-l-2)(\Delta+l-1)}g_{\Delta,l}-l(l+1)\frac{(\Delta-2)(\Delta-l)(\Delta+l-2)}{16\Delta((l-\Delta)^2-1)(\Delta+l-1)}g_{\Delta+2,l-2}\right)
	\end{dmath}
	 In our convention, the integrals involving $P1^2$ and $R1^2$ do not contain the factor of $l(l-1)$. The conformal integrations related to the tensor structures $P1 \times R1$ will vanish.

	\bibliographystyle{unsrt}
	\bibliography{references}
	
\end{document}